\documentclass{article}

\everydisplay{\mathsurround 0pt} 

\usepackage[affil-it]{authblk}
\usepackage{amsmath}
\usepackage{amssymb}
\usepackage{amsthm}
\usepackage{color}
\usepackage[a4paper]{geometry}
\usepackage[colorlinks,citecolor=blue,urlcolor=blue]{hyperref}
\usepackage{mathtools}
\usepackage[round]{natbib}
\usepackage[inline]{enumitem}
\usepackage{mathrsfs}

\usepackage{natbib}

\usepackage{algorithm}
\usepackage{algorithmic}

\usepackage{eucal} 
\usepackage{latexsym} 
\usepackage{accents}
\usepackage{subcaption}
\usepackage{cancel}
\usepackage{rotating}
\usepackage{xcolor}

\newcommand{\GG}[1]{}

\newcommand{\R}{\mathbb{R}}

\newtheorem{theorem}{Theorem}
\newtheorem{definition}{Definition}

\begin{document}

\title{The Wasserstein Impact Measure (WIM): a generally applicable, practical tool for quantifying prior impact in Bayesian statistics}

\author{Fatemeh Ghaderinezhad\thanks{Corresponding author: {fatemeh.ghaderinezhad@ugent.be}}}
\affil{Department of Applied Mathematics, Computer Science and Statistics, Ghent University, Ghent, Belgium}

\author{Christophe Ley}
\affil{Department of Applied Mathematics, Computer Science and Statistics, Ghent University, Ghent, Belgium}

\author{Ben Serrien
\affil{Experimental Anatomy Research Group, Vrije Universiteit Brussel, Brussels, Belgium}}

\maketitle

\begin{abstract}
The prior distribution is a crucial building block in Bayesian analysis, and its choice will impact the subsequent inference. It  is therefore important to have a convenient way to quantify this impact, as such a measure of prior impact will help us to choose between two or more priors in a given situation. A recently proposed approach consists in determining the Wasserstein distance between posteriors resulting from two distinct priors, revealing how close or distant they are. In particular, if one prior is the uniform/flat prior, this distance leads to a genuine measure of prior impact for the other prior. While highly appealing and successful from a theoretical viewpoint, this proposal suffers from severe practical limitations: it requires prior distributions to be nested, posterior distributions should not be of a too complex form, in most considered settings the exact distance was not computed but sharp upper and lower bounds were proposed, and the proposal so far is restricted to scalar parameter settings. In this paper, we overcome all these limitations by introducing a practical version of this theoretical approach, namely the Wasserstein Impact Measure (WIM). In three simulated scenarios, we will compare the WIM to the theoretical Wasserstein approach, as well as to two competitor prior impact measures from the literature. We finally illustrate the versatility of the WIM by applying it on two datasets.

\end{abstract}


\section{Introduction}

With the increase in computational power readily available in personal computers and the increase in complexity in statistical models, more and more researchers are shifting towards Bayesian statistics. Within statistics, research domains such as Bayesian nonparametrics, Bayesian learning or dynamic Bayesian networks have emerged and are flourishing, and in numerous other fields the Bayesian paradigm is strongly sought after. For instance, \cite{RBB2018} used Bayesian inference to identify parameters in viscoelasticity, hereby showing the substantial influence of the prior on viscoelasticity; \cite{KKW2006} consider a storm depth multiplier model for rainfall uncertainty, and use various priors for the multiplier variance to quantify this uncertainty; 
in clinical trials the incorporation of historical/prior information is especially relevant, for example in pediatric drug evaluation \citep{GSBZGG17}; and in the recent COVID-19 crisis, prior information learnt from one part of the world (typically the province Wuhan in China) represents invaluable information to predict the evolution in other countries with the help of the Bayesian framework \citep{BRV2020}.

The Bayesian framework allows more flexibility and more intuitive interpretations compared to
frequentist methods, but this same flexibility comes with the cost of argumentation for certain
choices and the assessment of their impact on any conclusions drawn from the data. In Bayesian
statistics, particularly the choice of the prior can have a profound impact on the inferences drawn
from data, especially at small sample sizes. The prior elicitation thus can be perceived as dual: an opportunity to introduce important prior knowledge into the problem, but at the same time a challenge when it comes to choosing the right prior.  The latter induces an essential research question: how can one quantify the impact of the choice of the prior on the posterior distribution, and hence on subsequent inference?  \cite{DiaFre1986a, DiaFre1986b} showed that under  certain regularity conditions  the effect of the prior wanes as the sample size increases, however in practice the sample size of course is always finite and, very often, rather small, underlining the relevance of said question. At first sight, a pragmatic answer could be a sensitivity analysis, but such an analysis often depends on how it is carried out and it only covers certain aspects of the inference, hence cannot provide a satisfying general answer. 

There does not exist a formal definition of prior impact in the literature. According to \cite{RMN2014}, an all-encompassing approach to this problem is seemingly philosophically and mathematically impossible. The mathematical problem lies in the lack of proper definition, while the philosophical problem concerns different schools of Bayesian inference: for a subjective Bayesian, the impact of the prior may be of less concern than for an objective Bayesian or frequentist statistician. Consequently, different measures of prior impact have been proposed over the years. A popular approach is the so-called \emph{effective sample size}, defined as the approximate number of observations equivalent to the information conveyed by the prior, see for instance \cite{LPC2007, Morita08, RMN2014, Wiesenfarth19, JTC2020} and the references therein.  \cite{Ker2011} introduced the Neutrality, corresponding to the posterior's tail probability to the left of the frequentist maximum likelihood estimate. Yet another approach has been taken by \cite{LRS2017a} who measure the Wasserstein distance between two posteriors, of which one results from the  prior of interest and the other is the no-prior data-only  posterior. Since it is mostly impossible to calculate this  distance explicitly, the authors have provided sharp lower and upper bounds on the Wasserstein distance and their approach relies on a variant of the famous Stein Method. In order to compare any two priors directly,  \cite{GhaLey2019a} recently extended their approach to any two priors for one dimensional-parameters, provided that the posteriors are nested; see also \cite{GhaLey20}.

For practical purposes, the power of the Wasserstein distance idea has not  been exploited so far. The obtained bounds and rare explicit results are obtained for  tractable posteriors; indeed, most examples considered in \cite{LRS2017a} and \cite{GhaLey2019a} are (related to) conjugate priors. Moreover, their results are confined to the priors whose supports are nested and to the one-dimensional setting, meaning that prior impact can only be assessed for one scalar parameter at once. Nice as they are, these mostly theoretical results thus are not yet broadly usable in practice. The aim of the present paper is to precisely fill this important gap and make this theoretically successful method widely usable in practice. More concretely, we will provide in Section~\ref{sec:WIM} the Wasserstein Impact Measure, abbreviated WIM, for assessing prior impact for any type of priors and any dimensions. The WIM relies on a numerical computation of Wasserstein distances and will allow us to compare any two priors, thus making the WIM a fully usable alternative to the proposals from the literature. By means of Monte Carlo simulations, we shall  show how  close the theoretical upper and lower bounds from \cite{LRS2017a} and \cite{GhaLey2019a} are to the computed WIM, hereby shedding new light on the tightness of these bounds. In Section~\ref{sec:compar}, we will then compare the WIM to other prior impact measures from the literature, namely the effective sample size (in what follows we will use its most recent version called MOPESS) and the Neutrality mentioned above. We illustrate the practical aspects of the WIM on two data examples in Section~\ref{sec:data}. Finally, we conclude the paper with a discussion in Section~\ref{sec:discu}.
\section{The WIM for  prior impact}\label{sec:WIM}
We will now present our new practice-oriented measure of the impact of the choice of the prior, the WIM. For a given dataset and, hence, a given likelihood, it quantifies the Wasserstein distance between two posteriors resulting from two distinct priors. In particular, if one prior is the (improper) uniform/flat prior, then this distance reveals how much information the other prior adds to the posterior in comparison to the likelihood alone. To properly introduce the WIM, we will first briefly describe the Wasserstein distance (Section~\ref{sec:Was}), followed by the prior impact measure of  \cite{LRS2017a} and \cite{GhaLey2019a} (Section~\ref{sec:Stein}). These preliminaries then allow us to describe the WIM  in Section~\ref{sec:VW}, along with its multi-dimensional extension. In order to see how the WIM compares to the upper and lower bounds on the theoretical Wasserstein distance provided in \cite{LRS2017a} and \cite{GhaLey2019a}, we will investigate three examples by means of Monte Carlo simulations in Section~\ref{sec:simus}.

\subsection{The Wasserstein distance}\label{sec:Was} 
The Wasserstein distance \citep{Vas1969}, also known as the Mallows distance \citep{Mal1972}, is a fundamental   metric  on the space of probability measures. It mostly is based on the theory of optimal transport (it has some different names in other fields of studies such as Monge-Kantorovich-Rubinstein distance in the physical sciences, the optimal transport distance in optimization, the earth mover's distance in computer science) and gives a natural measure of the distance between two distributions. A formal definition goes as follows.
\begin{definition}
Let $(\mathcal{X},d)$ be a complete metric space with metric $d: \mathcal{X} \times \mathcal{X} \longrightarrow \R^+$. The \textit{Wasserstein distance of order $p$ $(p \geq 0)$} between two Borel probability measures $\nu_1$ and $\nu_2$ on $\mathcal{X}$ is defined as
$$
d_{W_p}(\nu_1 , \nu_2) = \left\{ \underset{\nu \in \Pi(\nu_1,\nu_2)}\inf \int_{\mathcal{X} \times \mathcal{X}} (d(x,x'))^p \nu (dx,dx') \right\}^{1/p},
$$
where $\Pi(\nu_1,\nu_2)$ is the set of all Borel probability measures on $\mathcal{X} \times \mathcal{X}$ with marginals $\nu_1$ and~$\nu_2$. 
\end{definition}
According to  \cite{SoMu2017}, the following three characteristics make the Wasserstein distance particularly attractive in various applications and explain our choice for this metric:
\begin{itemize}
\item[$\bullet$] incorporation of a ground distance on the space in question that causes adequacy compared to competing metrics such as the Total Variation or $\chi^2$-metrics {which do not consider any similarity structure on the ground space};
\item[$\bullet$] clear and intuitive interpretation {as the amount of ``work'' required to turn one probability distribution into another}; 
\item[$\bullet$] well established capacity \citep{RTG2000} of   the Wasserstein distance to  capture human perception of similarity,  making it particularly fashionable in computer vision and related fields.
\end{itemize}
In \cite{LRS2017a} and \cite{GhaLey2019a},  the Wasserstein-1 distance has been adopted, which can be re-expressed as
$$
d_W(P_1,P_2) = \underset{h \in \mathcal{H}}{\sup}\, |{\rm E}[h(X_1)] - {\rm E}[h(X_2)]|,
$$
where $\mathcal{H}$ represents the class of Lipschitz-1 functions for random variables $X_1$ and $X_2$ with respective distribution functions $P_1$ and $P_2$.


\subsection{Quantifying prior impact via the Wasserstein distance}\label{sec:Stein}

We start by fixing the notations. Let $X_1,\ldots,X_n$ be a set of independent and identically distributed observations from a parametric model with parameter of interest $\theta \in \Theta \subseteq \R$. In our setting, the distribution of the $X_i$'s may be discrete or continuous. In addition, we simply denote the sampling distribution or likelihood function of the observations by $\ell (x;\theta)$ where $x=(x_1,\ldots,x_n)$ are the observed realizations of $X_1,\ldots,X_n$. Now consider two distinct (possibly improper)  prior densities $p_1(\theta)$ and $p_2(\theta)$ for our parameter of interest $\theta$. According to Bayes' Theorem, the two resulting posteriors for $\theta$ are 
$$
p_i(\theta;x) =\kappa_i(x) p_i(\theta) \ell(x;\theta),  \  \  i=1,2,
$$
where $\kappa_1(x)$ and $\kappa_2(x)$ are normalizing constants. We respectively denote by $(\Theta_1,P_1)$ and $(\Theta_2,P_2)$  the couples of random variables and cumulative distribution functions corresponding to the posterior densities $p_1(\theta;x)$ and $p_2(\theta;x)$. The Wasserstein distance $d_W(P_1,P_2)$ then reveals how close the posteriors are and, consequently, how similar the related inferences will be. This consideration lies at the heart of the approach of \cite{LRS2017a}, where one prior is the uniform/flat prior. When $n\rightarrow\infty$ the distance goes to 0 since the prior impact wanes, illustrating the theory of \cite{DiaFre1986a, DiaFre1986b}. At any finite sample size $n$, the distance thus shows the impact the prior has on the posterior for the given dataset $x$. However, calculating explicitly this distance is most often impossible to achieve, which is why sharp upper and lower bounds are needed. A crafty technique for reaching upper bounds is the so-called Stein Method, which we will not describe here since it is  not needed for our purposes. We refer the interested reader to  \cite{Ross} and \cite{LRS2017b}. For our bounds we however do need the concept of Stein kernel $\tau_i$ of $P_i$ which is defined as
$$
\tau_i (\theta ; x) = \frac{1}{p_i(\theta;x)} \int_{a_i}^{\theta} (\mu_i-y) p_i(y;x) dy, \  \  i=1,2,
$$
where $a_i$ is the lower bound of the support $I_i=(a_i,b_i)$ of $p_i(\cdot;x)$ and $\mu_i$ is the (assumed finite) mean of $\Theta_i$, $i=1,2$. This function is always positive and vanishes at the boundaries of the support. An important assumption of the framework of \cite{LRS2017a} and \cite{GhaLey2019a}   is that the posterior densities $p_1(\theta;x)$ and $p_2(\theta;x)$ are \textit{nested}, meaning that one support is included in the other. Supposing $I_2 \subseteq I_1$ allows us to write
$$
p_2(\theta;x) =\frac{\kappa_2(x)}{\kappa_1(x)} \rho(\theta) p_1(\theta;x)
$$
 with $\rho (\theta) = \frac{p_2(\theta)}{p_1(\theta)}$. For  two general priors $p_1(\theta)$ and $p_2(\theta)$, \cite{GhaLey2019a}  obtained the following result.
\begin{theorem}[Ghaderinezhad and Ley (2019)]\label{maintheo}
Consider $\mathcal{H}$ the set of Lipschitz-1 functions on $\R$. Assume that $\theta\mapsto\rho(\theta)$ is differentiable on ${I}_2$ and satisfies (i) ${\rm E}[(\Theta_1-\mu_1)\rho(\Theta_1)]<\infty$, (ii) $\left(\rho(\theta)\int_{a_1}^{\theta}(h(y)-{\rm E}[h(\Theta_1)])p_1(y;x)dy\right)'$ is integrable for all $h\in\mathcal{H}$ and (iii) $\lim_{\theta\rightarrow a_2,b_2}\rho(\theta)\int_{a_1}^{\theta}(h(y)-{\rm E}[h(\Theta_1)])p_1(y;x)dy=0$ for all $h\in\mathcal{H}$. Then
\begin{equation}\label{bounds}
|\mu_1-\mu_2|=\frac{|{\rm E}[\tau_1(\Theta_1;x)\rho'(\Theta_1)]|}{{\rm E}[\rho(\Theta_1)]}\leq d_{W}(P_1,P_2)\leq \frac{{\rm E}[\tau_1(\Theta_1;x)|\rho'(\Theta_1)|]}{{\rm E}[\rho(\Theta_1)]}
\end{equation}
and, if the variance of $\Theta_1$ exists,
\begin{equation*}
|\mu_1-\mu_2|\leq d_{W}(P_1,P_2)\leq ||\rho'||_{\infty}\frac{{\rm Var}[\Theta_1]}{{\rm E}[\rho(\Theta_1)]}
\end{equation*}
where $||\cdot||_{\infty}$ stands for the infinity norm.
\end{theorem}
The three regularity conditions in Theorem~\ref{maintheo} are mild assumptions and are nearly always satisfied in practical situations. By choosing the improper prior $p_1(\theta)=1$, the prior impact measure of \cite{LRS2017a} is retrieved. The fact that the same quantities appear in the upper and lower bounds gives a hint as to why the bounds in~\eqref{bounds} are sharp.  \cite{GhaLey2019a} moreover showed that if, in addition to the conditions of Theorem~\ref{maintheo}, we assume that the ratio $\rho$ is monotone increasing or decreasing, then
$$
d_{W}(P_1,P_2)= \frac{{\rm E}[\tau_1(\Theta_1;x)|\rho'(\Theta_1)|]}{{\rm E}[\rho(\Theta_1)]}.
$$
We illustrate these theoretical bounds and exact distances on a few examples in Section~\ref{sec:simus}. This measure of prior impact is highly intuitive and theoretically appealing, but suffers from drawbacks: the method relies on the assumption of nested densities, more complicated posterior structures do not lend themselves to calculable bounds, the extension to the case of multi-dimensional parameters (either two or more parameters associated with  scalar observations or a parameter vector associated with multi-dimensional observations) is a highly non-trivial task due, \emph{inter alia}, to the not yet well developed multi-dimensional Stein's Method (see \cite{MRS20}), and, even if the bounds are well computable, they may not give us accurate information on the actual distance if they are spread far apart. In order to solve all these drawbacks and propose a more practically usable version of this intuitive type of prior impact, we give the Wasserstein Impact Measure  in the next section.

\subsection{The computable Wasserstein Impact Measure (WIM)}\label{sec:VW}

In order to overcome the drawbacks of the rather theoretical approach presented in the previous section whilst keeping its advantages, and in order to build a practically usable impact measure, we use the Vallender formula \citep{Val1974}. This formula allows one to calculate the Wasserstein distance by using  the cumulative distribution functions $F_i(\theta;x)$ of the two posterior distributions $P_1$ and $P_2$ according to the formula
\begin{equation}\label{VW1}
d_W (P_1 , P_2) = \int_a^b |F_1 (\theta;x) - F_2 (\theta;x)| d\theta,
\end{equation}
where $a$ and $b$ are the bounds of the support of the parameter of interest (support that obviously can be $\R$). This is an exact expression, and numerical integration techniques permit to calculate this quantity irrespective of any assumption of nested supports. Now, it is not uncommon to encounter complicated posterior densities for which the cdfs are not computable. In such cases, we suggest to have recourse to computational techniques such as Markov Chain Monte Carlo (MCMC) methods to generate random samples from each distribution $P_1$ and $P_2$ and then use Monte Carlo integration. For instance, the \textit{"transport"} package in $\mathtt{R}$ offers functions for estimating the Wasserstein distance of any order between two sets of samples from different distributions, see  \cite{Schetal2019}. The transport package solves optimality problems by computing the Wasserstein distance and  includes functions to obtain numerical computations of $p$-th order Wasserstein distances. Most of the functions in the package have been designed for  data with two or higher dimensions. Thus, this approach not only solves the problem of dealing with complicated posteriors but also the multi-dimensional issue. Indeed,~\eqref{VW1} can readily be extended to $m>1$ parameters $\theta_1,\ldots,\theta_m$ as follows:
$$
d_W (P_1 , P_2) = \int_{a_1}^{b_1}\ldots \int_{a_m}^{b_m} |F_1 (\theta_1,\ldots,\theta_m;x) - F_2 (\theta_1,\ldots,\theta_m;x)| d\theta_1\cdots d\theta_m
$$
where $a_j,b_j$ are the bounds of the support of $\theta_j$, $j=1,\ldots,m$. 

It is this computable Wasserstein impact measure that we term WIM. To quantify its uncertainty for a given dataset, we suggest having recourse to bootstrapping. To illustrate how the WIM works and, in particular, how it compares to the theoretical upper and lower bounds given in the previous section, we will study three concrete examples in the next section. 

\subsection{Practical illustration and comparison of the WIM with the theoretical upper and lower bounds}\label{sec:simus}
A natural question of interest is to find out how close our computational estimate of the Wasserstein distance lies to the upper and lower bounds provided analytically in examples considered by \cite{LRS2017a} and \cite{GhaLey2019a}. This not only yields a comparison of both prior impact measures, but can also be of independent interest to  probabilists working with the Wasserstein distance and, in particular, researchers from the Stein Method community. The three models we consider are the binomial, the Poisson and the normal distributions. For the Poisson example, the Wasserstein distance happens to yield an exact distance, which thus allows us to check how close our computational  estimate is to the true Wasserstein distance.

In each of the  examples, our strategy is as follows. We first present the theoretical upper and lower bounds on the Wasserstein distance (or, in case of the Poisson, the exact distance), and then for various choices of parameters and/or sample sizes we generate random samples  from the model under investigation. For each sample, we then calculate the bounds on the Wasserstein distance between two posteriors as well as our WIM.

\subsubsection{Priors for the Poisson model}
The most famous count distribution is the Poisson with probability mass function
$$
x\mapsto \frac{\exp(-\theta) \theta^{x}}{x!}
$$
where $\theta \in \R^+$ is the parameter of interest indicating the average number of events in a given time interval, and $x \in N$ is the number of occurrences. The Gamma distribution is one of the most popular choices of priors for the Poisson model, in large parts due to the fact that it is a conjugate prior. For example, it was used to study asthma mortality rates by  \cite{Geletal2004}. The Gamma probability density function  is 
$$
\theta\mapsto \frac{\beta^\alpha}{\Gamma(\alpha)} \theta^{\alpha-1} \exp(-\beta \theta) 
$$
where $\alpha, \beta > 0$.  The Gamma prior contains as special cases  the exponential prior $(\alpha = 1)$, the uniform prior $(\alpha = 1, \beta = 0)$, and the Jeffreys prior which is proportional to $\theta^{-1/2}$ $(\alpha = 1/2 , \beta = 0)$. In addition,  \cite{Ker2011} showed that the $Gamma(1/3, 0)$ prior is  optimal in terms of Neutrality for a large range of sample sizes and values of $\theta$. Neutral priors in this context lead to posterior distributions with approximately $50$ percent probability to be either smaller or larger than the maximum likelihood estimate, see Section~\ref{sec: neutral}. \cite{GhaLey2019a} presented the exact Wasserstein distance when the ratio of  two $Gamma(\alpha_i,\beta_i), i=1,2,$ priors is monotone (corresponding to either $(1): \alpha_1 < \alpha_2 \cap \beta_1 > \beta_2$ or $(2): \alpha_1 > \alpha_2 \cap \beta_1 < \beta_2$):
\begin{equation}\label{Poidist}
d_W (P_1, P_2) = \frac{1}{n+\beta_1} \left| \alpha_2 - \alpha_1 - (\beta_2 - \beta_1) \frac{\alpha_2 + \sum_{i=1}^n x_i}{n+\beta_2} \right|,
\end{equation}
where $x_i, i=1,\ldots,n,$ are the data. Interestingly, during exploratory calculations through simulations we found out that this exact distance was also very accurate for parameter combinations not meeting criteria (1) or (2). 

\begin{figure} 
\centering
\includegraphics[width=\linewidth]{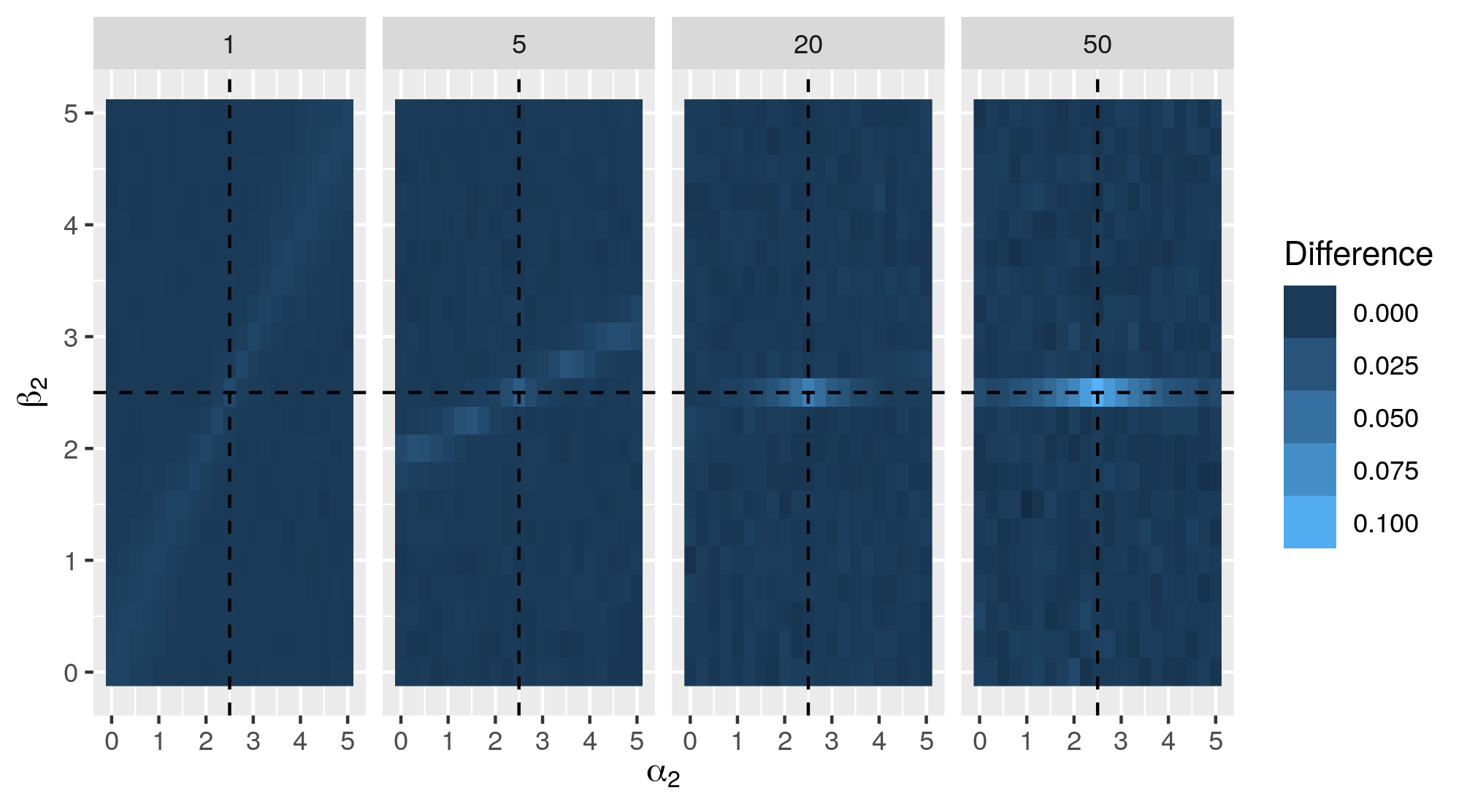}
  \caption{Difference between the theoretical Wasserstein distance and our computational estimate of the Wasserstein distance between posteriors resulting from a $Gamma(2.5,2.5)$ and a $Gamma(\alpha_2,\beta_2)$ prior. The columns indicate different values of $\theta$, and the sample size is fixed throughout at $n=10$. The vertical and horizontal dashed lines indicate the values 2.5 of $\alpha_1$ and $\beta_1$, respectively. The lower right and upper left quadrants in each panel are combinations of Gamma parameters where expression~\eqref{Poidist} holds.}\label{Fig34}
\end{figure}

For our Monte Carlo simulation study, we then consider $Gamma(2.5,2.5)$ as a fixed  prior of reference for different combinations of $\theta$ and $(\alpha_2,\beta_2)$ as second $Gamma(\alpha_2,\beta_2)$ prior. Under the aforementioned conditions we draw 1000 random samples of size $n = 10$ for the values $\theta=1, 5, 20$ and 50. For each pair of posteriors, we draw 10000 random samples from them and calculate the Wasserstein distance between them with the \textit{transport} package, as explained in Section~\ref{sec:VW}. Figure \ref{Fig34}  shows the results of all combinations of the Gamma parameters, including cases where criteria (1) or (2) are not met. The differences  between the true Wasserstein distance and the computational estimate are very low, indicating that the WIM is accurate throughout. Over all values of $\theta$ and $(\alpha_2,\beta_2)$ the median difference between both values was 0.00078 (corresponding to $0.08\%$ of the true value). For  sample sizes larger than $n=10$, the computational estimate  lies even closer to the true Wasserstein distance.  This shows that, at least for the Poisson model with Gamma priors, the WIM is highly accurate, validating thus our computational approach. 

\subsubsection{Priors for the success parameter in a binomial model}

\begin{figure} 
\centering
\includegraphics[width=\linewidth]{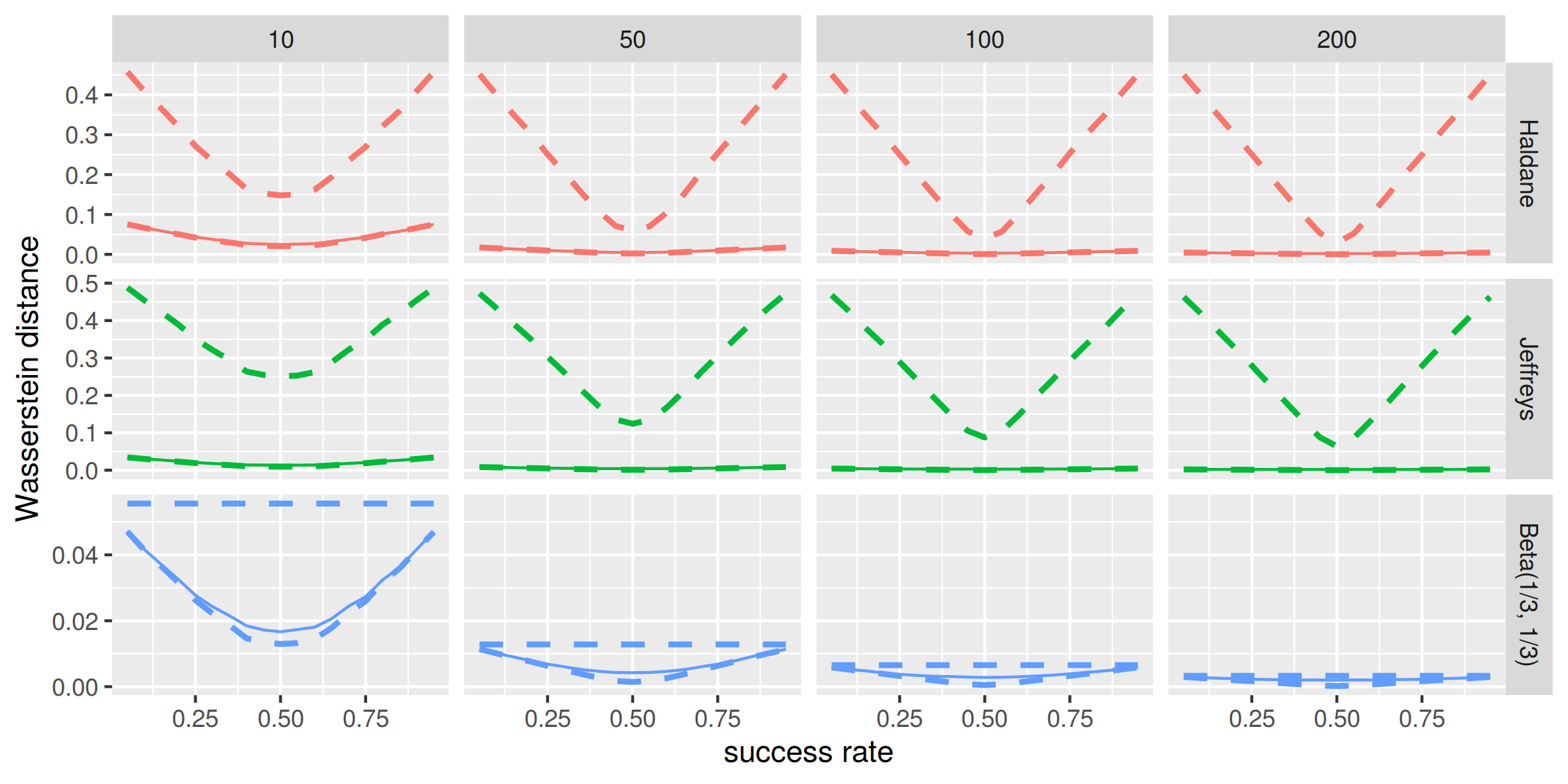}
  \caption{Comparison of the theoretical upper and lower bounds on the Wasserstein distance between posteriors  (dashed lines) with the WIM (full line) for different sample sizes ($n = 10, 50, 100, 200$; columns) and success rate ($\theta = 0.05, 0.10, \cdots, 0.95$; x-axis). Each prior (rows) is compared to the uniform prior. Note the different scales for the y-axes in each panel: the $Beta(1/3,1/3)$ prior is not only neutral, but also very close to the uniform prior (low impact).}\label{Fig31}
\end{figure}

Let us now consider the  {binomial} distribution $Bin(n, \theta)$  with probability mass function
$$
x\mapsto {{n} \choose {x}} \theta^x (1-\theta)^{n-x}
$$
where $x \in \{0, \ldots, n\}$ is the number of observed successes, the natural number $n$ indicates the number of binary experiments and $\theta \in (0,1)$ stands for the success parameter. The conjugate prior for the binomial distribution is the beta model $Beta(\alpha, \beta)$ with parameters $\alpha,\beta >0$. The $Beta(\alpha,\beta)$ prior contains some specific cases based on the definition of $\alpha$ and $\beta$, such as the uniform prior $(\alpha = \beta =1)$, the Jeffreys prior $(\alpha = \beta = 1/2)$, Haldane's prior $(\alpha = \beta =0)$ which gives a complete uncertainty, and the neutral prior $(\alpha = \beta =1/3)$ \citep{Ker2011}. From Theorem~\ref{maintheo} (whose assumptions are  satisfied) straightforward calculations yield the following upper and lower bounds (see \cite{LRS2017a} and \cite{GhaLey2019a}  for details):
\let\labelitemi\labelitemii
\begin{itemize}
\item[$\bullet$] $Beta(\alpha,\beta)$ versus uniform:
\begin{eqnarray*}
\left| \frac{x+1}{n+2}\left(\frac{\alpha+\beta-2}{n+\alpha+\beta}\right) - \frac{\alpha-1}{n+\alpha+\beta} \right| &\leq& d_W(P_1,P_2) \\
&\leq& \frac{1}{n+2} \left( |\alpha-1|+\frac{x+\alpha}{n+\alpha+\beta} (|\beta-1|-|\alpha-1|)  \right);
\end{eqnarray*}
\item[$\bullet$] Jeffreys versus uniform:
\begin{eqnarray*}
\frac{|\frac{n}{2}-x|}{(n+2)(n+1)} \leq d_W(P_1,P_2) \leq \frac{1}{n+2}\left(\sqrt{\frac{(x+1/2)(n-x+1/2)}{(n+2)(n+1)^2}} + \left| \frac{x+1/2}{n+1} - \frac{1}{2} \right|\right);
\end{eqnarray*}
\item[$\bullet$] Haldane versus uniform:
\begin{eqnarray*}
\frac{2|\frac{n}{2}-x|}{n(n+2)} \leq d_W(P_1,P_2) \leq \frac{2}{n+2}\left( \sqrt{\frac{x(n-x)}{n^2(n+1)}} + \left| \frac{x}{n} - \frac{1}{2} \right|\right).
\end{eqnarray*}
\end{itemize}
For various combinations of the parameters $n$ and $\theta$, we draw 1000 random binomial observations and calculate the bounds above. Concerning the WIM,  for each pair of posteriors, we draw 10.000 random samples from them and calculate the Wasserstein distance between them with the \textit{transport} package. The results are indicated in Figure~(\ref{Fig31}). Not unexpectedly, the highest accuracy happens around the center of the parameter space and it decreases towards the edges for all priors and sample sizes. We see that the WIM actually lies very close to the lower bound for all priors, sample sizes and success rates, meaning that the upper bound is likely to be an overly conservative estimate of the true Wasserstein distance compared to the lower bound. This sheds an insightful new light on the theoretical bounds. Also note that upper and lower bounds are very close for the neutral $Beta(1/3,1/3)$ prior.

\subsubsection{Priors for the normal model}
\begin{figure} 
\centering
\includegraphics[width=\linewidth]{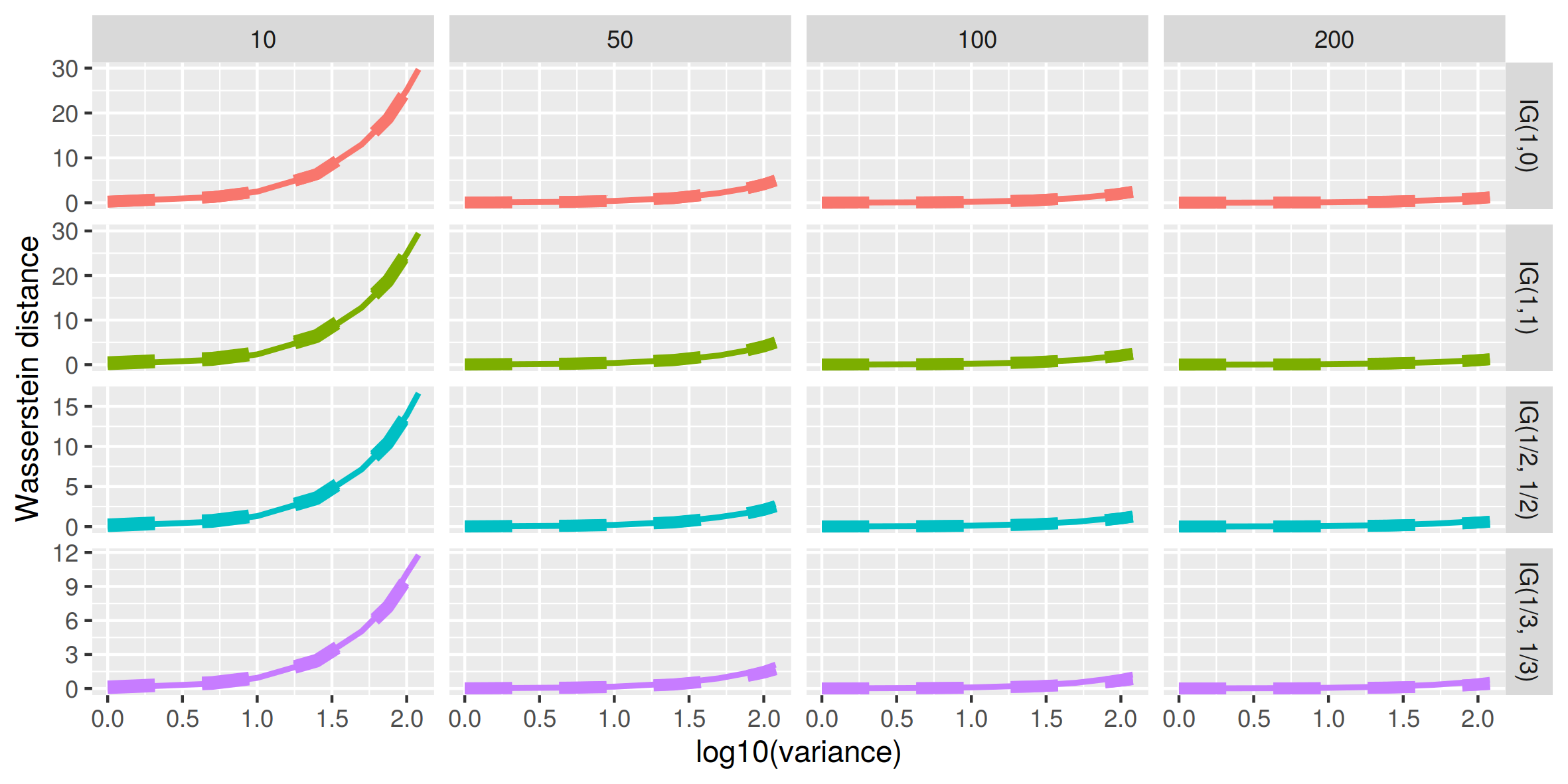}
  \caption{Wasserstein distance between posteriors based on different Inverse Gamma priors (rows) vs Jeffreys' prior for different sample sizes (columns) and values of $\sigma^2$ (x-axis). The dashed lines represent the theoretical upper and lower bounds and the full line the WIM. The values of $\sigma^2$ (0.5, 1, 5, 10, 25, 50, 75, 100, 120) were plotted on a log-10 scale which gave a slightly better view, but still the bounds are barely distinguishable.}\label{Fig37}
\end{figure}

Finally we choose as continuous model the normal distribution with probability density function 
$$
x\mapsto \frac{1}{\sqrt{2\pi} \sigma} \exp \left(-\frac{(x-\mu)^2}{2\sigma^2} \right),  \  \  x \in \R,
$$
with location parameter $\mu \in \R$ and dispersion parameter $\sigma > 0$. We consider $\sigma^2$ to be the parameter of interest and $\mu$ is fixed (known a priori). We wish to compare here the improper Jeffreys prior $p(\sigma^2) \propto 1/\sigma^2$, invariant under reparameterization, and the Inverse Gamma (IG) prior with positive real parameters $\alpha$ and $\beta$ which happens to be the conjugate prior for the normal distribution.  \cite{GhaLey2019a} have shown that the Wasserstein distance between posteriors derived under the two aforementioned priors is bounded as
$$
\frac{|\frac{\alpha}{2} \sum_{i=1}^n (x_i-\mu)^2 - (\frac{n}{2} - 1)\beta|}{(\frac{n}{2}+ \alpha - 1)(\frac{n}{2} -1)} \leq d_W(P_1,P_2) \leq \frac{\left( \frac{\alpha}{2} \sum_{i=1}^n (x_i-\mu)^2 + \frac{n\beta}{2}+\beta(2\alpha-1) \right)}{(\frac{n}{2}+ \alpha - 1)(\frac{n}{2} -1)},
$$
where $x_i, i=1,\ldots,n$ are the  data. We inspect four IG priors including $IG(1,0)$ (uniform prior on the positive real line), $IG(1,1)$, $IG(1/2,1/2)$ and $IG(1/3,1/3)$. Fixing $\mu$ at 0, we again draw, for  different values of $\sigma^2$ and $n$,  1000 random normal datasets of size $n$,  and take the average over the 1000 draws of both the upper and lower bounds and the WIM, calculated in the same way as in the previous examples. The results are shown in Figure~\ref{Fig37}.  The bounds are very narrow over the range of examined variances, sample sizes and priors. 
We can conclude that there is hardly a noticeable difference between the upper/lower bounds and the WIM for the normal model with IG priors for the squared scale parameter. 

\section{Comparison with prior impact measures from the literature}\label{sec:compar}

After comparison of the WIM with the theoretical upper and lower bounds on the Wasserstein distance from \cite{LRS2017a} and \cite{GhaLey2019a}, we now compare the WIM to other competitors from the literature, namely the Neutrality  and the MOPESS (abbreviation to be defined below). We start by describing both approaches in Sections~\ref{sec: neutral} and~\ref{sec:MOPESS}, respectively, before proceeding to the comparison with our WIM in Section~\ref{sec:compare}.
\subsection{Neutrality}\label{sec: neutral}
 \cite{Ker2011} introduced the concept of \textit{Neutrality} $N$ for a given prior with  corresponding posterior $\Theta$, say. The Neutrality of $\Theta$ is defined as the probability of $\Theta$ to lie on the left of the frequentist maximum likelihood estimate $\hat{\theta}_{MLE}$, which in mathematical terms means
$$
N = P(\Theta < \hat{\theta}_{MLE}) = \int_{a}^{\hat{\theta}_{MLE}} p (\theta;x) d\theta,
$$
where $p(\theta;x)$ is the posterior and $a$ the lower bound of its support. The closer this tail probability is to 1/2, the less informative or the more neutral the prior is. \cite{Ker2011} showed that for the binomial and Poisson likelihoods, the conjugate $Beta(1/3,1/3)$ and  $Gamma(1/3,0)$ are respectively the most neutral priors over the entire parameter space. The advantage of the Neutrality $N$ as a metric of prior impact is that it is an \textit{absolute} metric for each prior, in contrast to the WIM and the MOPESS (see next section) which are relative measures.  Another advantage is the ease of calculation in conjugate models, since it can be calculated analytically and, when MCMC has been used, it can be readily calculated based on the posterior samples. In addition, the scale of $N$ is the same for all models no matter how complex they are since $N \in [0,1]$.  However, it cannot be used properly in cases where the frequentist MLE is at the boundaries of the parameter space which is a notable disadvantage in Bayesian analysis. For example, in a binomial model when the MLE is $0$ or $1$, then $N$ will be the same for any prior. Moreover,  \cite{Ker2011} has not mentioned how to extend this concept to  multivariate or multiparameter situations, so we do not discuss this issue here.

\subsection{Mean Observed Prior Effective  Sample Size (MOPESS)}\label{sec:MOPESS}
In certain studies one distinguishes between the effect of the prior and the amount of information the prior does contain.  For conjugate models, the nominal amount of information in the prior is known and is often expressed in terms of the number of pseudo-observations (prior sample size (PSS)). For instance, in a beta-binomial model, the parameters of the $Beta(\alpha,\beta)$ prior can be interpreted as the number of successes $(\alpha)$ and failures $(\beta)$ in the available prior information. When the prior and the likelihood function differ substantially or are very comparable, the impact will be different although the prior has the same amount of information in each situation. Therefore the concept of effective prior sample size (EPSS) has been introduced.

We will follow here the definition of \cite{RMN2014} of a {new class of effective prior sample size measures based on prior-likelihood discordance}, which is also referred to as (the degree of) prior-likelihood conflict in \cite{JTC2020}. \cite{RMN2014} published the first algorithm on how to adjust for prior-likelihood conflict in the calculation of EPSS to answer the question: how many extra observations are needed to transform a posterior based on a baseline prior into the posterior based on the prior of interest? The EPSS can be lower or higher than the nominal PSS (information) depending on the actual data that are observed. When the prior mean is arbitrarily far from the maximum likelihood estimate, then the impact will become larger. The EPSS can also be negative, indicating that the baseline prior is in fact more impactful than the prior of interest. \cite{JTC2020} have extended the method of  \cite{RMN2014} to a more general setting and their method also works for lower sample sizes (where the impact of the prior is particularly important). They calculate the mean observed prior effective sample size (MOPESS) according to the following steps:
\begin{enumerate}
\item Derive the posteriors based on the prior of interest and the baseline prior.
\item Use the posterior predictive distribution based on the prior of interest to sample two sets of $m$ additional observations ($m=1,\ldots,L$ where $L$ is the maximum feasible value for EPSS).
\item For each $m$, calculate 
\begin{enumerate}[label=(\roman*)] 
\item the Wasserstein-2 distance between posteriors based on (1) original data combined with prior of interest versus (2) original $+$ additional data combined with baseline prior $(W1)$.
\item the Wasserstein-2 distance between posteriors based on (3) original data combined with baseline prior versus (4) original + additional data combined with prior of interest $(W2)$.
\end{enumerate}
\item The $m$ for which this distance is smallest is the OPESS. The lowest value of the set of $W1$ and $W2$ determines the sign: when $W1$ contains the lowest distance, then the prior of interest is more impactful than the baseline prior (OPESS $> 0$) and vice versa (OPESS $< 0$).
\item This process is repeated several times and the mean of the OPESS values is the MOPESS.
\end{enumerate}
An advantage of the MOPESS algorithm is that it indicates which of the two priors has the most impact thanks to the added sign. Similarly to our Wasserstein approach, the aim of the MOPESS is to compare any two posteriors and measure the relative impact of the priors. Hence, it is necessary to label one of the priors as the baseline. A   practical disadvantage of this method is the computation time for non-conjugate models where advanced sampling algorithms are necessary and need to be repeated several times. A further disadvantage of the MOPESS appears in  models with covariates such as in regression settings, because it requires assumptions to be made on the distribution of each covariate.  \cite{JTC2020} mention this issue, and also show through a simple linear regression example that the MOPESS in principle should work in higher dimensions, though this extension has not been touched upon outside of this one example.

\subsection{Comparison with the WIM}\label{sec:compare}

We will now revisit the three examples considered in Section~\ref{sec:simus} and study in how far our WIM relates to the Neutrality and the MOPESS. In order to see whether they provide similar information or we can  learn different aspects from them, we plot the WIM versus both the Neutrality $N$ and the MOPESS for the various priors corresponding to the Poisson, binomial and normal settings. The prior impact measures are obtained along the same lines as described in Section~\ref{sec:simus}. 

\subsubsection{Poisson case}

\begin{figure}
\centering
\includegraphics[width=\linewidth]{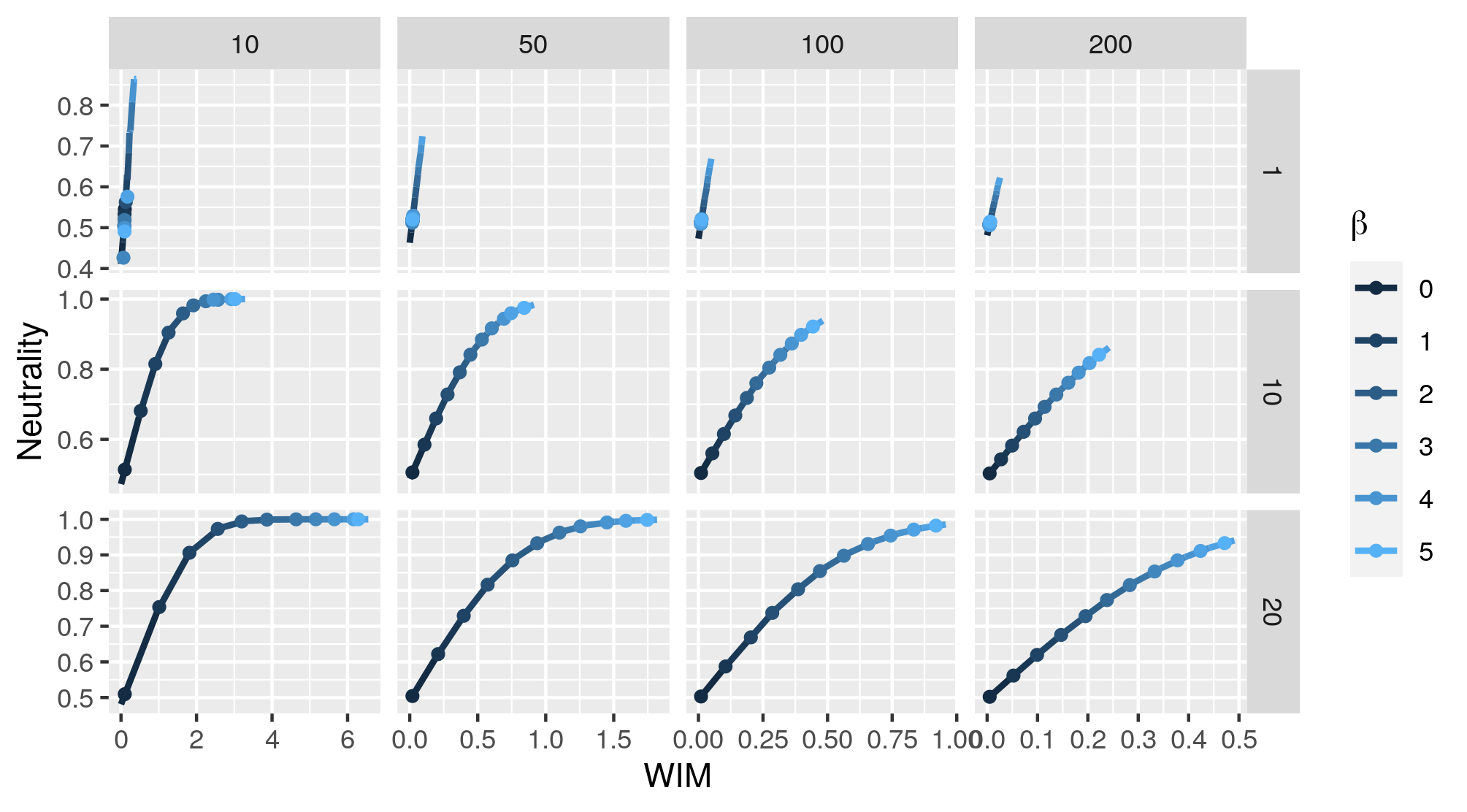}
  \caption{WIM \mbox{vs.} Neutrality  for the Poisson model. The columns correspond to different sample sizes and the rows to different values of $\theta$. The $\beta$ parameter of the priors is shown on the colour scale. The full line represents the impact measures for the $Gamma(1,\beta)$ \mbox{vs.} uniform prior and the dots for the $Gamma(\beta,\beta)$ \mbox{vs.} Uniform prior.} \label{Fig35}
\end{figure}

\begin{figure}
\centering
\includegraphics[width=\linewidth]{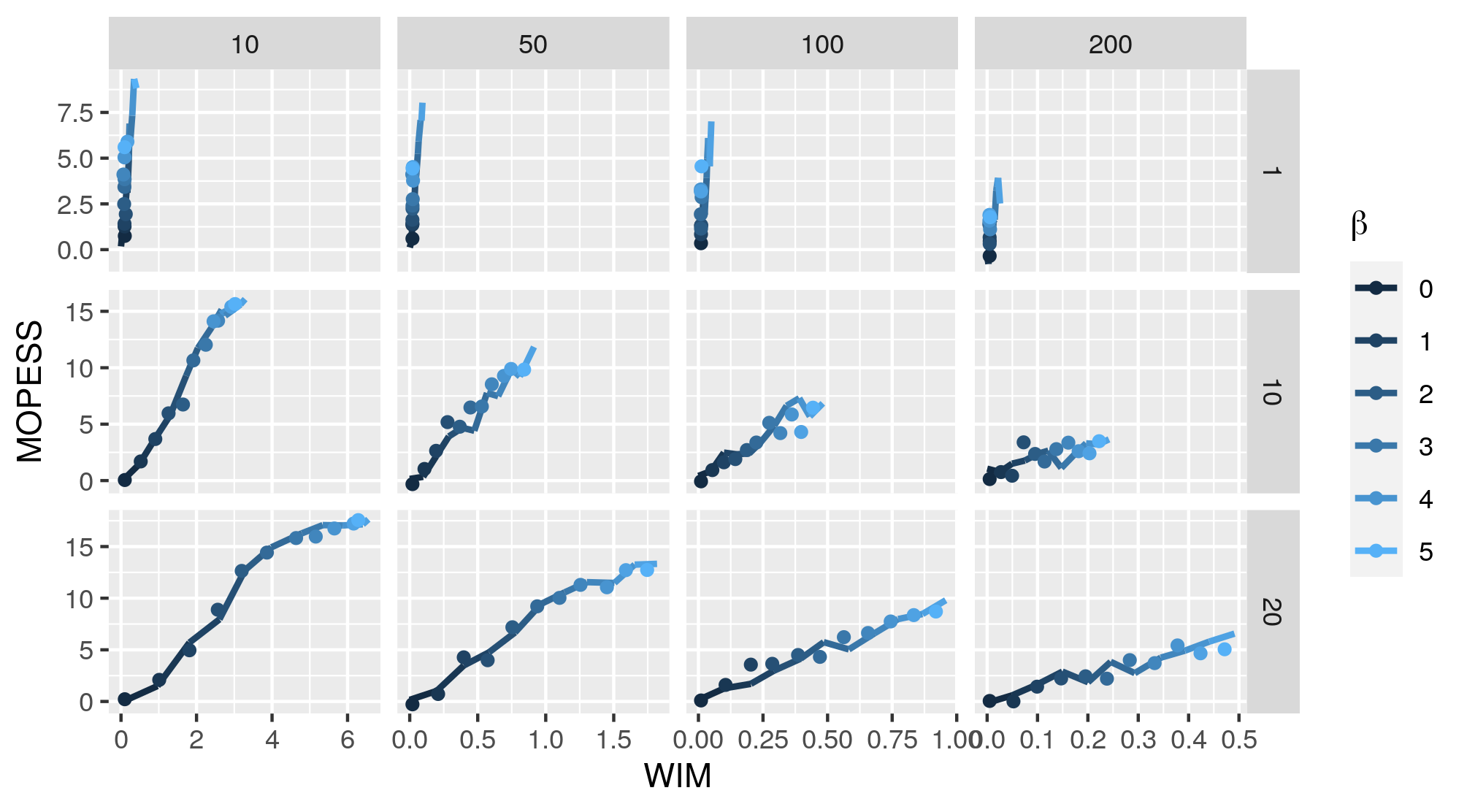}
  \caption{WIM \mbox{vs.} MOPESS  for the Poisson model. The columns correspond to different sample sizes and the rows to different values of $\theta$. The $\beta$ parameter of the priors is shown on the colour scale. The full line represents the impact measures for the $Gamma(1,\beta)$ \mbox{vs.} uniform prior and the dots for the $Gamma(\beta,\beta)$ \mbox{vs.} uniform prior.} \label{Fig36}
\end{figure}

Figures~\ref{Fig35} and~\ref{Fig36} show the WIM (x-axis) versus the Neutrality and the MOPESS (y-axis), respectively, for different priors and sample sizes in the Poisson case. For the WIM and the MOPESS, the  baseline prior is the uniform, which is not required for the Neutrality $N$. We observe that all prior impact measures  decrease with sample size, as one could expect.

The first clear conclusion for Neutrality against WIM is that these measures are very alike for the two considered Gamma priors, except for $n=10$. This is remarkable because the shapes of these priors diverge considerably with increasing $\beta$. Moreover, in some panels the Neutrality quickly reaches its maximum value meaning the Neutrality is no longer sensitive to distinguish between priors. In such cases, the WIM is a much  better choice to quantify the impacts of the priors as it is unbounded. Also in some cases such as $\theta = 1$ the WIM is not much affected by $\beta$.

The relation between the MOPESS and the WIM is  not  monotone in some panels due to the known high variability of the MOPESS \citep{JTC2020}, which speaks in favour of the WIM. One can notice the higher MOPESS values compared to the WIM, in particular for $\theta=1$ there is more variability in the MOPESS. Its positive sign  indicates that the Gamma priors are more impactful than the baseline uniform prior, an information  the WIM does not yield.


\subsubsection{Binomial case}

\begin{figure} 
\centering
\includegraphics[width=\linewidth]{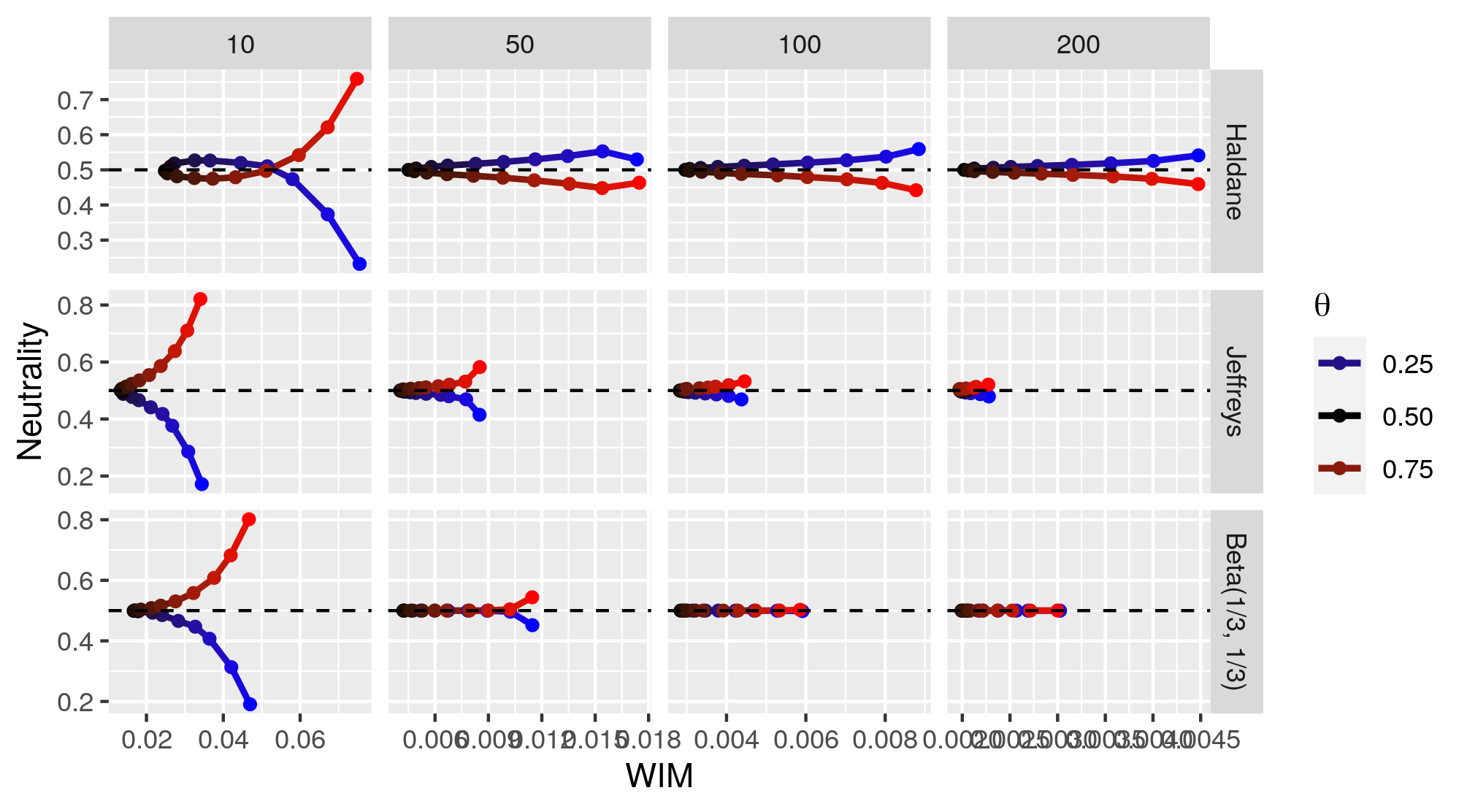}
  \caption{WIM \mbox{vs.} Neutrality  for the binomial model for different priors (rows) and sample sizes (columns). The color scale presents the binomial success rate $\theta = 0.05, 0.10, \cdots, 0.95$. The baseline prior is the uniform. Note the different axis scales.}\label{Fig32}
\end{figure}

\begin{figure}
\centering
\includegraphics[width=\linewidth]{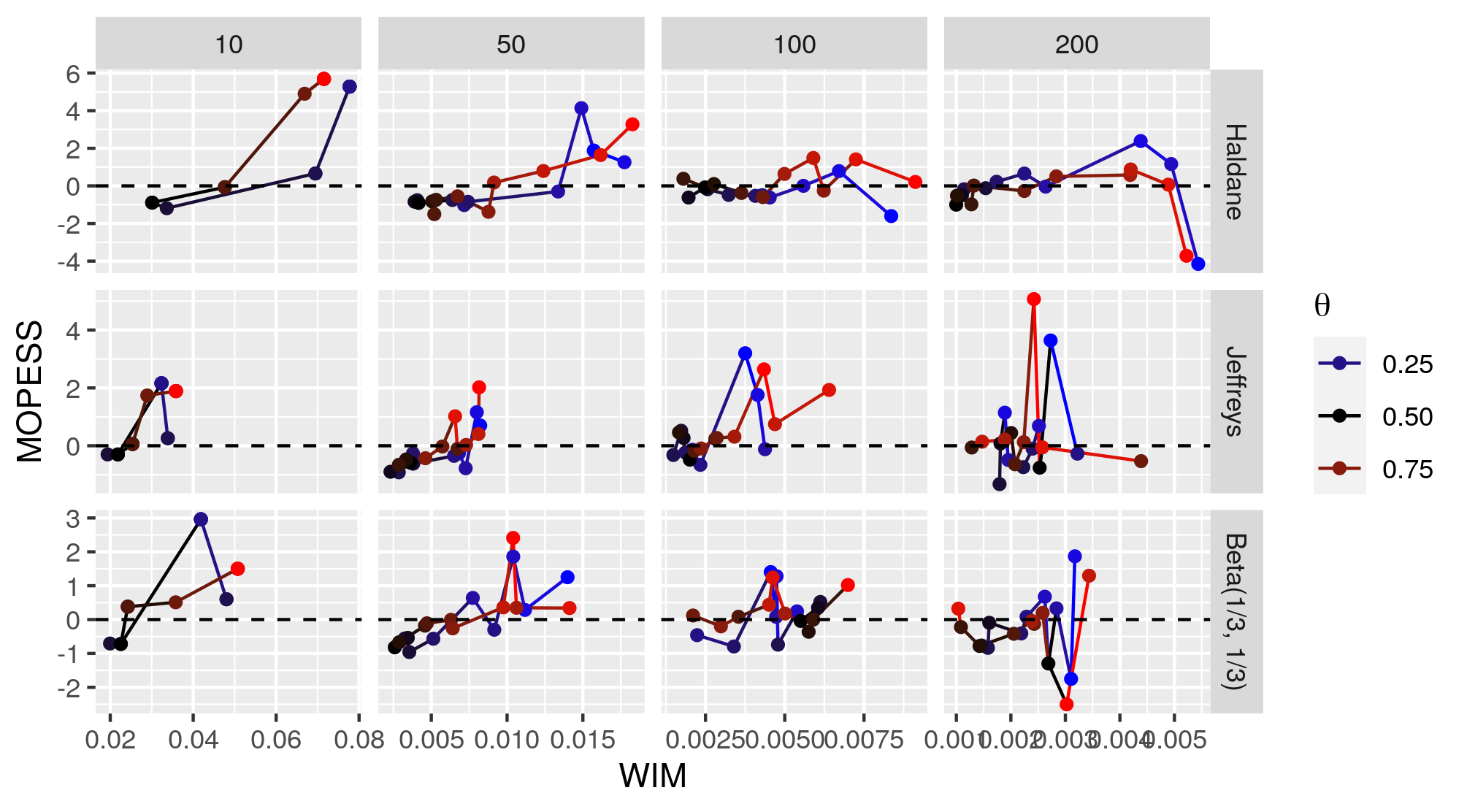}
  \caption{WIM \mbox{vs.} MOPESS  for the binomial model for different priors (rows) and sample sizes (columns). The color scale presents the binomial success rate ($\theta = 0.05, 0.10, \cdots, 0.95$). The baseline prior is the uniform. Note the different axis scales.} \label{Fig33}
\end{figure}

Figures~\ref{Fig32} and~\ref{Fig33} portray the WIM (x-axis) versus the Neutrality and the MOPESS (y-axis), respectively, for different priors and sample sizes in the binomial case.  For the WIM and the MOPESS, the  baseline prior is the uniform, which is not required for the Neutrality $N$. 

The optimal Neutrality of 1/2 is obtained for values of $\theta$ near 1/2, just at the point where the WIM also reports the smallest impact. The further we move towards the edges of the parameter space, the higher the WIM between priors and the less Neutrality the priors have. For Jeffreys' prior, this relation is monotone for all sample sizes. For Haldane's prior the relation is monotone only for $n = 100, 200$; for $n = 50$ we see a small non-monotonicity near the outer edge and for $n = 10$, we see that the Neutrality reverses in sign with respect to 0.50 around $\theta = 0.20$ and 0.80. The WIM is here a better interpretable, because monotone prior measure when moving away in either sense from $\theta=0.50$. For the neutral $Beta (1/3,1/3)$ prior, the graph shows that it optimally preserves Neutrality near 1/2 for $n = 50, 100, 200$, where one thus resorts to the WIM to detect differences. Again both prior impact measures decrease with the sample size. 

Since the MOPESS is a highly variable metric, the relationship with the WIM is not nicely monotone as with the Neutrality (even after averaging over 1000 replicates). For Haldane's prior, the MOPESS lies close to zero except for $n = 10$ and $\theta = 0.95$. In addition, for the other combinations of sample size and success rate  we do observe an increase in the MOPESS towards the edges of the parameter space which makes sense as the Haldane prior inflates towards 0 and 1. A similar behaviour exists for the neutral prior and Jeffreys' prior for the sample size $n=10$ (increasing of the MOPESS near the edge of the parameter space), however for larger sample sizes, the relationship between the WIM, the MOPESS and the success rate is very erratic due to the MOPESS' variability. We also observe that the MOPESS does not decrease with sample size; the clearer structure of the WIM makes it a better understandable measure for the binomial case.


\subsubsection{The normal case}
\begin{figure} 
\centering
\includegraphics[width=\linewidth]{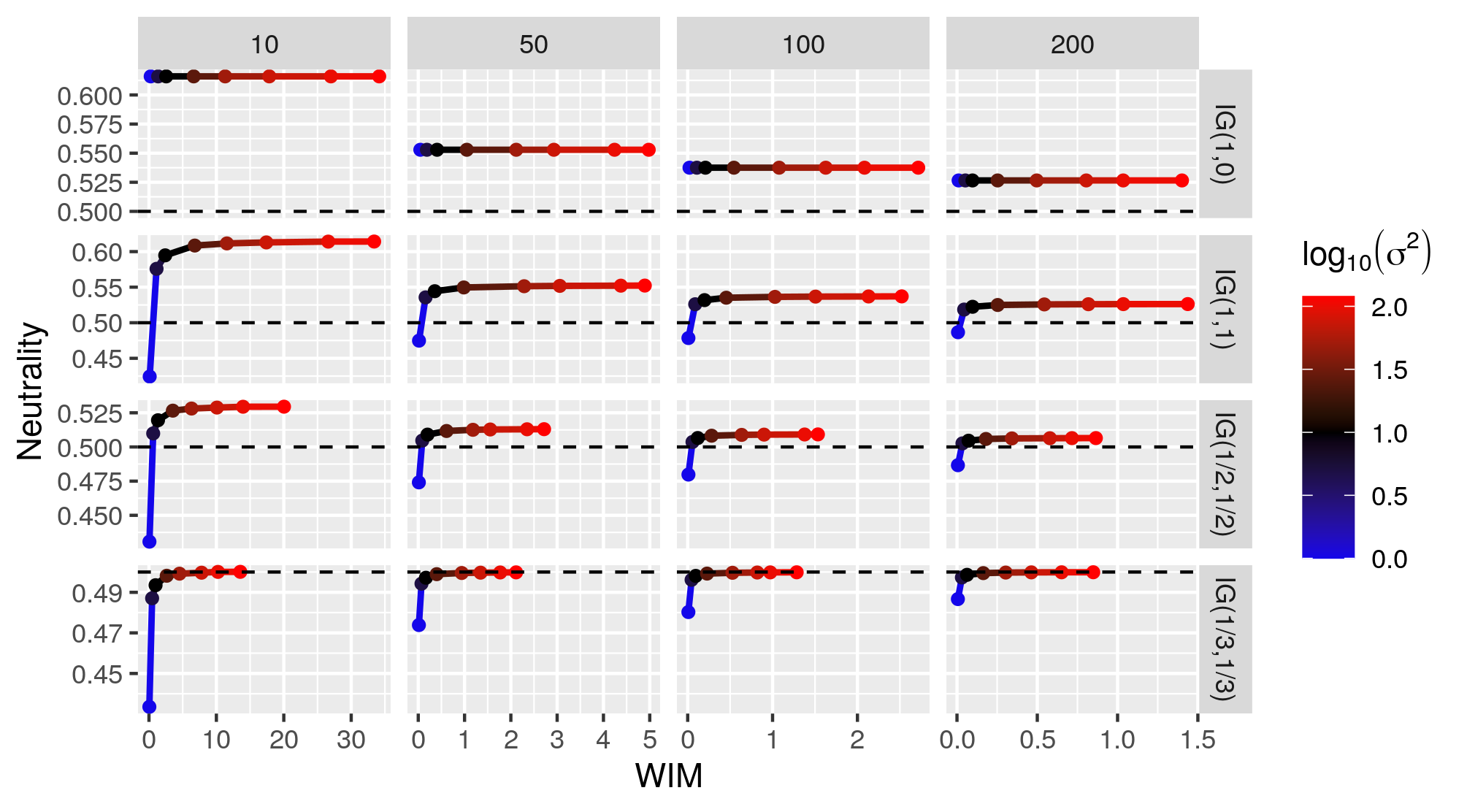}
  \caption{WIM \mbox{vs.} Neutrality  for the normal model for different  priors (rows)   and sample sizes (columns). The colour scale presents the log$_{10}$(variance). The baseline prior is the Jeffreys prior. Note the different axis scales.}\label{Fig38}
\end{figure}

\begin{figure} 
\centering
\includegraphics[width=\linewidth]{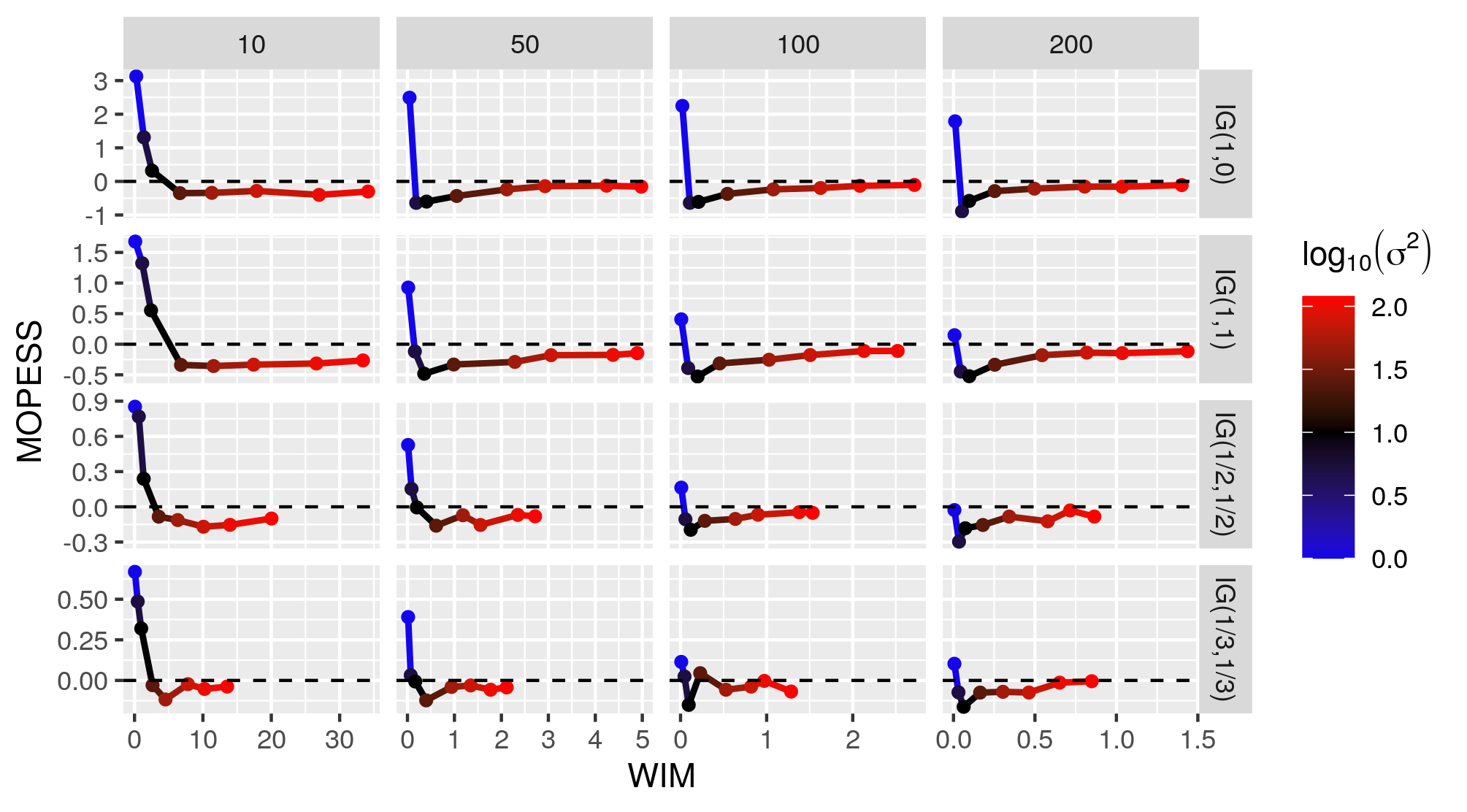}
  \caption{WIM \mbox{vs.} MOPESS  for the normal model for different  priors (rows)   and sample sizes (columns). The colour scale presents the log$_{10}$(variance). The baseline prior is the Jeffreys prior. Note the different axis scales.}\label{Fig39}
\end{figure}

Figures~\ref{Fig38} and~\ref{Fig39} present the WIM (x-axis) versus  the Neutrality and the MOPESS (y-axis), respectively, for different priors and sample sizes in the normal case. For the WIM and the MOPESS, the baseline prior is the Jeffreys prior, which is not required for the Neutrality $N$. All prior impact measures decrease with the sample size.

We first notice that the uniform $IG(1,0)$ prior's Neutrality does not depend on the variance and converges slowly to 1/2 as the sample size is increasing. Thus, for the comparison of the $IG(1,0)$ prior with Jeffreys' prior, the WIM is a more informative choice. For the other priors, the relationship is monotone at first until it reaches an asymptote for the Neutrality which is dependent on the sample size. For the $IG(1/3, 1/3)$ prior, we observe that this asymptote is  at 1/2 and reached rather quickly, indicating that Neutrality of this prior is optimal for a large range of values and the $IG(1/3, 1/3)$ thus can be considered a {neutral prior} as defined by  \cite{Ker2011}. It is notable that while the Neutrality converges to an asymptote and is thus no longer sensitive to distinguish between priors, the unbounded WIM is still able to distinguish further between the two priors.
 
For smaller values of $\sigma^2$, the MOPESS values are positive for all priors and sample sizes, which was expected because the priors under consideration have the dominant part of their probability mass near the limit of zero. So when the prior and likelihood are strongly similar, this corresponds to additional observations. However, only for the $IG(1,0)$ and $IG(1,1)$ priors is the MOPESS larger than 1. The MOPESS shrinks to zero or slightly below with increasing values of $\sigma^2$, indicating more impact from the Jeffreys prior. Although the relationship between the WIM and the MOPESS is less erratic than for the binomial model, the relationship is still not monotone. A possible reason for this behaviour is that the Wasserstein distance has no sign to indicate which of the two priors is more impactful. A second reason is that finite sampling variability limits the display of a monotone relationship \citep{JTC2020}. Both the MOPESS and WIM report insightful information and it seems hard to pick a better choice in this setting.
 
\section{Illustration of the WIM  on two data sets}\label{sec:data}
\subsection{Priors for the skewness parameter of the skew-normal  model}
\begin{figure} 
\centering
\includegraphics[width=\linewidth]{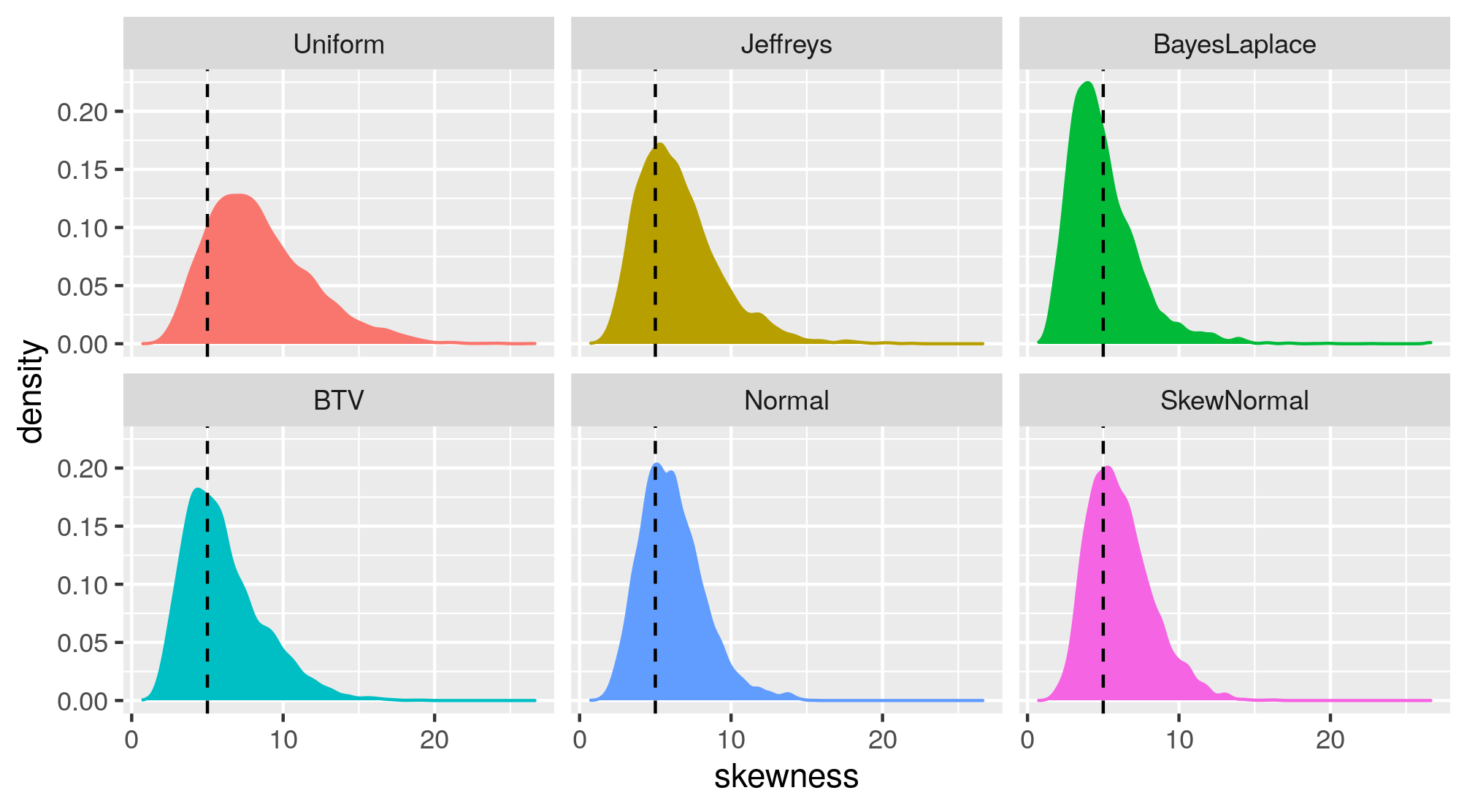}
  \caption{Posterior distributions for the skewness parameter of the skew-normal model for the frontier dataset based on different priors.}\label{Fig310}
\end{figure}

Many distributions have been built to capture skewness in data. Arguably the most famous instance is the skew-normal distribution of \cite{Azz1985} with probability density function 
\begin{eqnarray}\label{skewmodel}
x\mapsto  \frac{2}{\sigma} \phi \left(\frac{x-\mu}{\sigma}\right) \Phi \left(\alpha \frac{(x-\mu)}{\sigma} \right), \  \  \  x \in \R,
\end{eqnarray}
where $\mu \in \R$ is the location parameter, $\sigma \in \R^+$ the scale parameter, both inherited from the standard normal distribution with pdf denoted by $\phi$ and  cumulative distribution function $\Phi$, and $\alpha \in \R$ is called the skewness parameter. The density (\ref{skewmodel}) is an asymmetric model for $\alpha \neq 0$ and reduces to  a standard normal  when $\alpha=0$. A well-known problem of the skew-normal model in relation with frequentist inference is that, for some datasets, the maximum likelihood estimate of the skewness parameter becomes infinite \citep{AzzaCapit}. A famous example of such a situation  is the \textit{frontier} dataset, to be found in the \textit{sn} package of $\mathtt{R}$ where the MLE for the skewness is 1.4e+06 for 50 draws from a skew-normal with  $\mu=0,\sigma=1$ and skewness $\alpha=5$. In these cases a Bayesian estimation procedure seems a meaningful alternative. A review of different priors for the skew-normal model and more general skew-symmetric distributions is given in  \cite{GLL2020}.  We will apply various priors discussed in that paper to the frontier dataset and calculate the WIM  between the resulting posteriors.   The following priors will be examined (the * indicates that some priors are approximated by a known parametric form for ease of computation):
\begin{itemize}
\item[$\bullet$] Uniform/flat prior: $p(\alpha) \propto 1$. 
\item[$\bullet$] Jeffreys' prior (*): a tractable approximation of the Jeffreys prior $p(\alpha)$ is $ t_{0, \pi^2/4,0.5}(\alpha)$ where $t_{a,b,c}$ is the density of the Student $t$-distribution with location $a\in\R$, scale $b>0$  and $c>0$ degrees of freedom \citep{BB07}. This prior is proper, symmetric around zero, decreasing in $|\alpha|$ and its tails are of the order $O(\alpha^{-3/2})$.
\item[$\bullet$] Bayes-Laplace prior: following the Bayes-Laplace rule, \cite{BB07} proposed a uniform prior on the interval $[-1,1]$ for $\frac{\alpha}{\sqrt{1-\alpha^2}}$, which corresponds, for $\alpha$, to the prior $p(\alpha) = t_{0,0.5,2}(\alpha)$.
\item[$\bullet$] Beta Total Variation (BTV) prior (*): belonging to the class of distance-based priors proposed by \cite{DLR18}, the rationale for this prior on the skewness is that $\alpha$ not only controls the skewness, but shifts the entire distribution and hence a prior should rather be set on the Wasserstein distance between the normal and the skew-normal and, from there, one can derive a prior for $\alpha$. We choose the so-called $BTV(1,1)$ prior which can be approximated by $p(\alpha) = t_{0,0.92,1}(\alpha)$.
\item[$\bullet$] Normal prior:  an informative prior, chosen such that the mean is  zero in order to be comparable to the location of the other priors, and the variance is set to cover a reasonable scope of values, resulting in the prior ${\mathcal N}(0,5)$ \citep{CS13}.
\item[$\bullet$] Skew-normal prior: another informative  prior suggested by \cite{CS13}. It combines the location and scale of the ${\mathcal N}(0,5)$ with a skewness value of 2. 
\end{itemize}
All posterior distributions are numerically sampled by Markov Chain Monte Carlo with the T-walk algorithm in $\mathtt{R}$ \citep{ChrFox2010}. Based on  preliminary analyses, we used  chains of length 100k with a 50k burn-in and thinning rate of 5 as this yields stable results for this dataset. Figure~\ref{Fig310} gives an overview of the posteriors obtained from the different aforementioned priors. We  clearly observe that the mode of each posterior is very close to the true value $\alpha = 5$ of the frontier data. 

\begin{table}
\begin{center}
\begin{tabular}{|c c c c c c|}
\hline
Priors & Jeffreys & Bayes-Laplace & BTV & Normal & SkewNormal \\
\hline
Uniform & 1.604 & 3.138 & 2.213 & 2.081 & 2.067\\
Jeffreys &             &1.534 & 0.609  & 0.504 & 0.515 \\
Bayes-Laplace&    &          &  0.926 & 1.072 & 1.077\\
BTV                  &     &       &   &  0.452 &  0.453\\
Normal     &     &      &      &       & 0.040\\
\hline
\end{tabular}
\caption{WIM between pairs of posteriors resulting from different priors for the skewness parameter $\alpha$ in the skew-normal model for the frontier dataset.}\label{summary}
\end{center}
\end{table}

\begin{figure} 
\centering
\includegraphics[width=\linewidth]{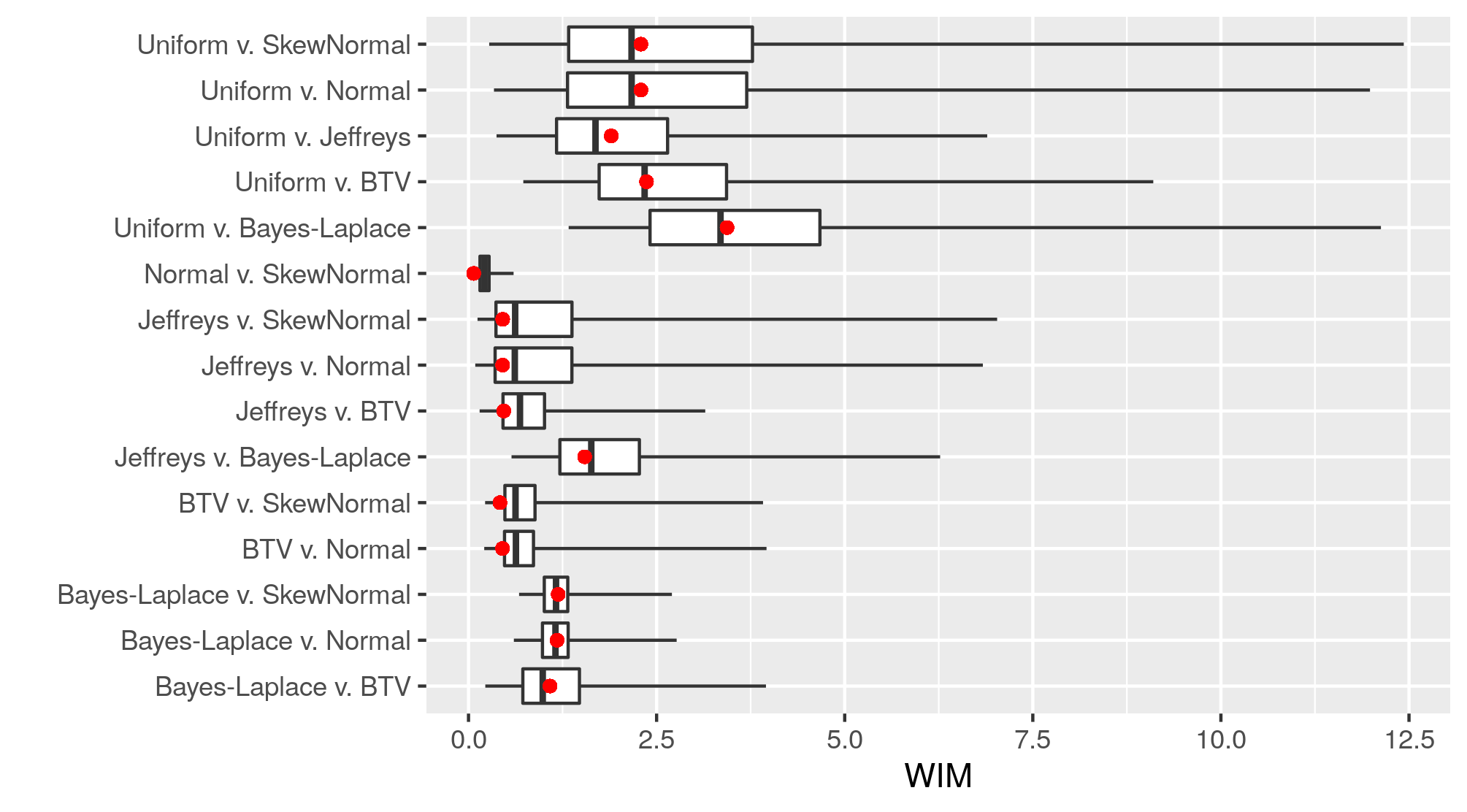}
  \caption{Boxplot of the WIMs between all pairs of posteriors resulting from different priors for the skewness parameter in the skew-normal model. The boxplots are based on $B=250$ bootstrap simulations and the red dots indicate the value of the WIM of the original "frontier" dataset.}\label{Fig311}
\end{figure}

Table~\ref{summary} contains the WIM for each pair of posteriors resulting from different priors. It reflects what can be seen in Figure~\ref{Fig310}, namely that the posterior based on the uniform/flat prior is considerably different from all the others. Setting the uniform prior as baseline prior, the Jeffreys prior would have the least impact compared to the other priors. The smallest distance occurs between the informative normal and skew-normal priors, indicating that it makes little difference which of these two we choose. It is interesting to note that the $BTV(1,1)$ prior, considered to be non-informative \citep{DLR18},  seems to have slightly bigger impact than these informative priors when compared to the prior-free case (uniform). This requires further investigation, which is why we also quantify the uncertainty behind these distances. To this end  we use bootstrap resampling (250 bootstrap samples) to obtain the sampling distribution of the WIM for all pairs of posteriors by calculating for each bootstrap sample of the {frontier} dataset the distance between the posteriors. The results are summarized  in Figure~\ref{Fig311}. The WIMs involving the uniform prior are not only the largest, but also have the highest variability. In particular, the WIM between the uniform and the informative priors has the largest variability, clearly larger than with the $BTV(1,1)$ prior, which is now well in line with the non-informative/informative character of these priors. Not surprisingly, the least variable distance occurs between the normal and the skew-normal priors. 

%

\subsection{Logistic regression and weakly informative priors}

\begin{figure}
\centering
\includegraphics[width=0.5\linewidth]{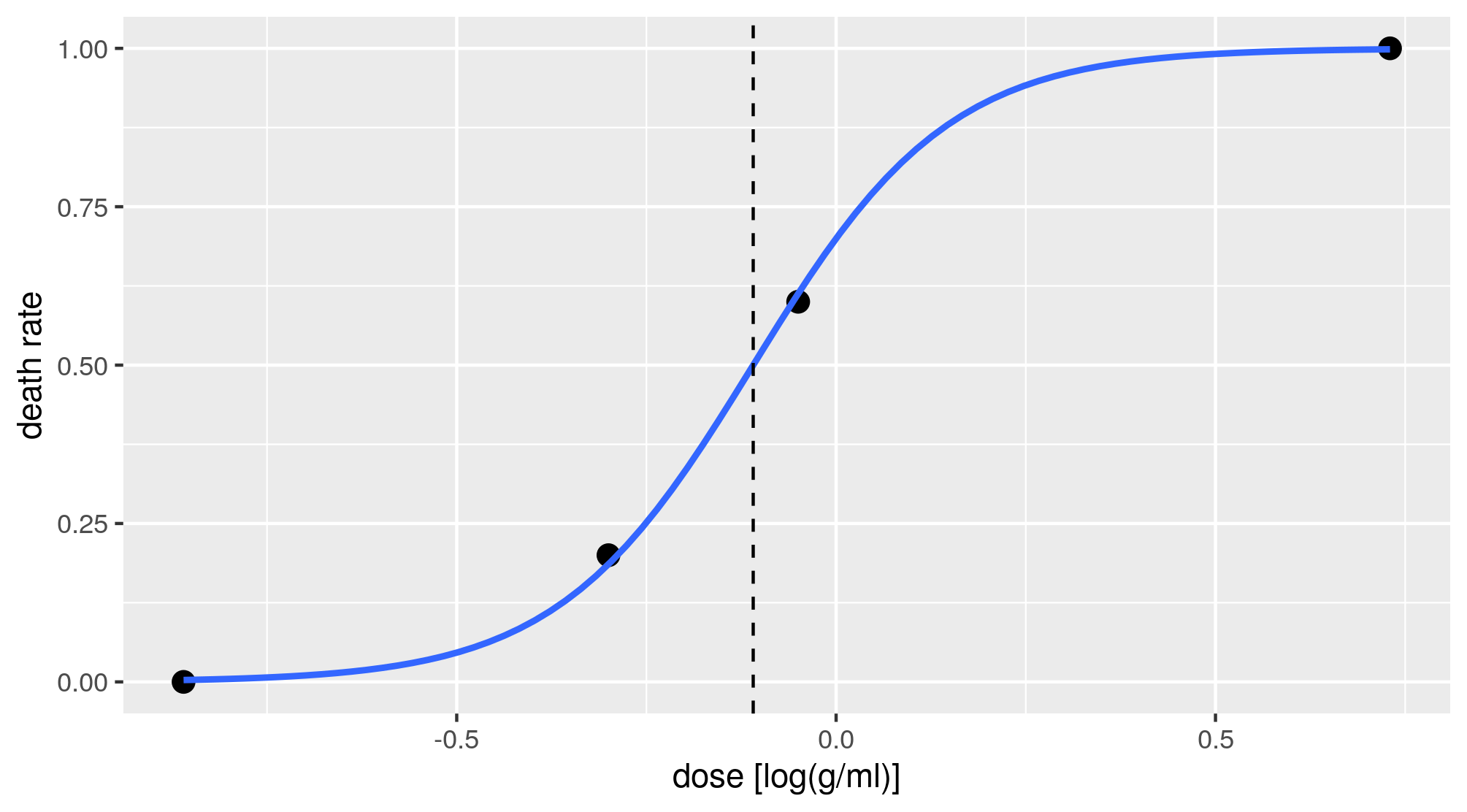}
  \caption{Bioassay experiment of 4 binomial observations (each based on 5 replicates). The blue line presents the logistic regression model using MLE. The vertical dashed line presents the LD50 estimate based on the MLE of the regression coefficients.} \label{Fig313}
\end{figure}

In the present section, we compare the uniform versus so-called \textit{weakly informative} priors for the logistic regression model involving a single continuous covariate:
$$
logit(\pi_i) = \beta_0 + \beta_1 x_i,
$$
where $\beta_0,\beta_1\in\R$ are the regression parameters, $\pi_i$ is the probability that observation $i$ is a success and the \textit{logit} function is the logarithm of the odds. An important application of Bayesian inference occurs in dose-response studies where the parameter of interest is the "LD50" (lethal dose), the dose $(x)$ where the probability of death $(\pi)$ is exactly $50\%$. Using maximum likelihood estimation, we can get a point estimate for this parameter $LD50 = -\frac{\beta_0}{\beta_1}$, however there is no standard solution for the standard error. Instead  Bayesian inference  allows deriving  the posterior for LD50  from the posterior samples of $\beta_0$ and $\beta_1$ \citep{Geletal2004}. Here we examine different priors for the data of  \cite{RGFS1986} that are reported in \cite{Geletal2008}. This is a small-scale bioassay experiment with only $n = 4$ binomial observations (each based on 5 replicates) for different doses. With small samples,  the choice of the prior is even more important than for larger sample sizes. In Figure~\ref{Fig313} the data are shown together with the fit obtained by maximum likelihood estimation which is nearly perfect. Two of the four observations are at the outer limit $(0,1)$ and exert much influence on the analysis. 

\begin{figure} 
\centering
\includegraphics[width=\linewidth]{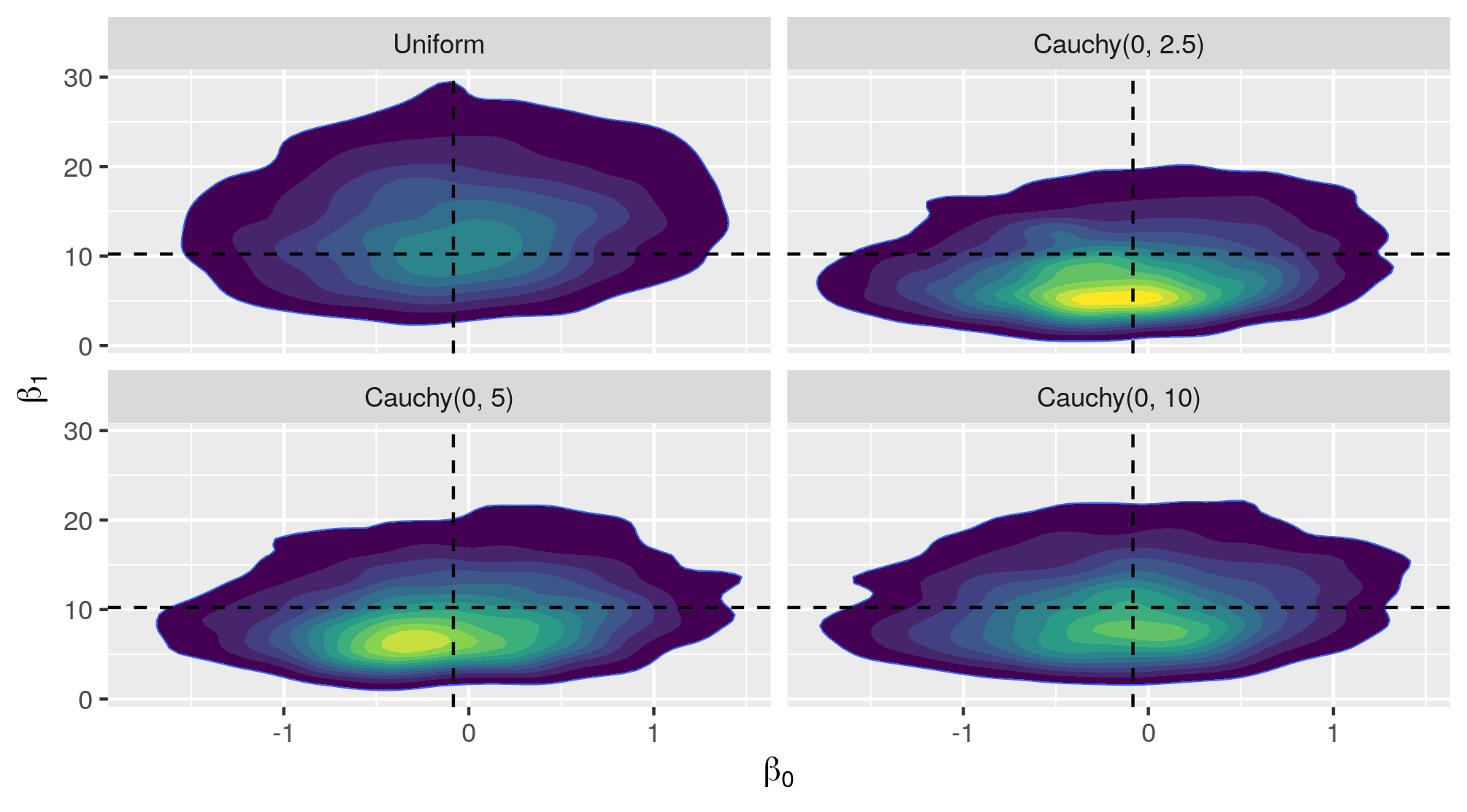}
  \caption{Joint posterior densities for $\beta_0$ and $\beta_1$ (the covariate was scaled according to \cite{Geletal2008}). Only the prior for $\beta_1$ is indicated in the Cauchy settings, since for $\beta_0$ we always use Cauchy(0,10). The dashed lines present the MLE of the regression coefficients. The colour scale reflects posterior density (dark blue = low density, yellow = high density).}\label{Fig314}
\end{figure}

\begin{figure} 
\centering
\includegraphics[width=\linewidth]{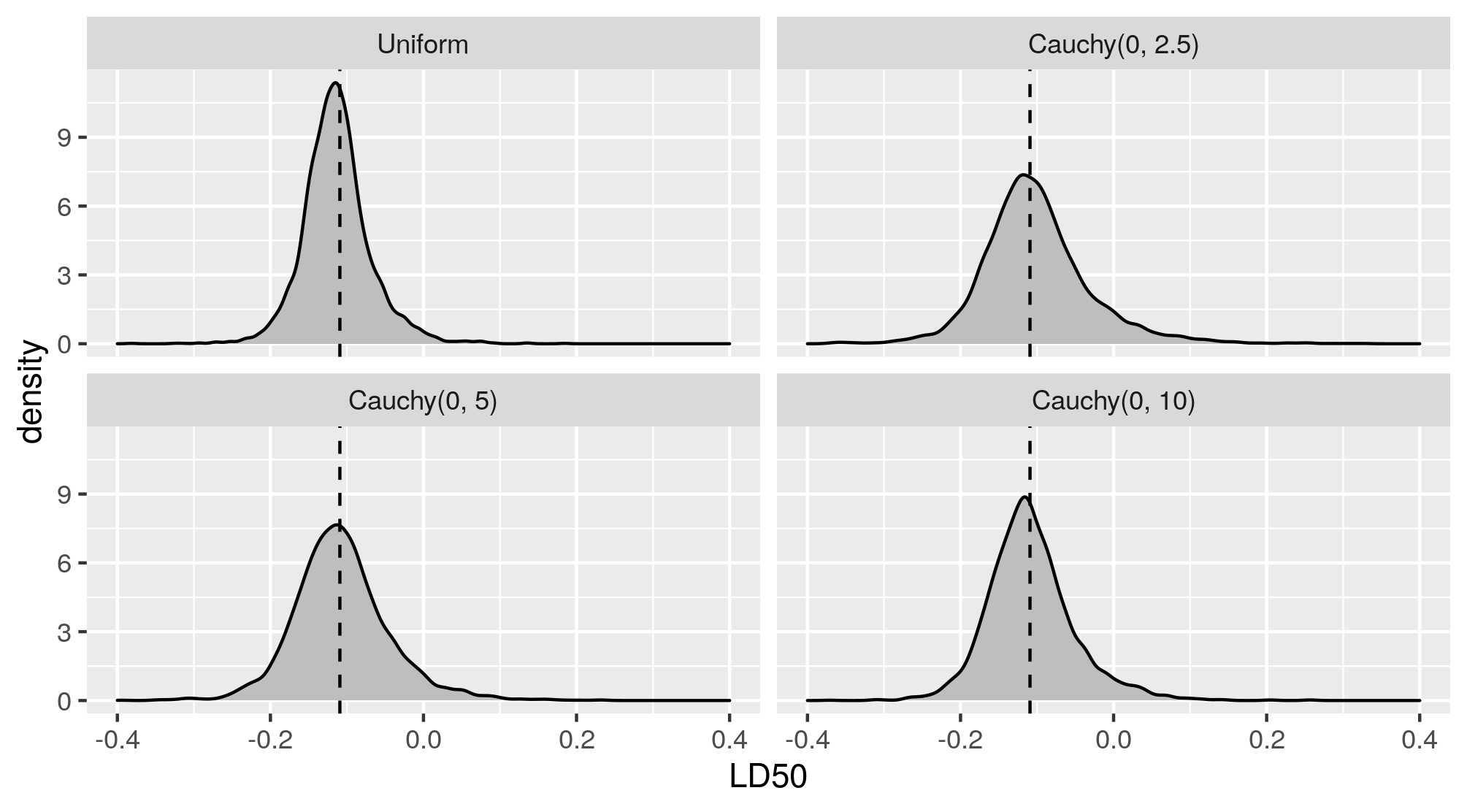}
  \caption{Posterior densities for the LD50 (lethal dose at which the probability of death is $50\%$). Only the prior for $\beta_1$ is indicated in the Cauchy settings, since for $\beta_0$ we always use Cauchy(0,10). The dashed lines present the point estimate based on the MLE of the regression coefficients.}\label{Fig315}
\end{figure}

The difference of this application with previous models is that we are now in a multivariate situation. There are two regression parameters that are of interest plus a derived parameter, LD50. This means that the measures of prior impact should be inspected both on the joint posterior distribution and on the marginals, as they could potentially yield different information on prior impact.  The two priors we will consider are the following:
\begin{itemize}
\item[$\bullet$] Uniform prior: The uniform prior places a uniform distribution on $\R^2: p(\beta_0,\beta_1) \propto 1$. 
\item[$\bullet$] Cauchy prior: \cite{Geletal2008} propose to use the Cauchy prior centered at zero with scale parameters 2.5 and 10 as default priors for the slope and intercept parameters, respectively. Before using these priors, all covariates (in this case only the dose) need to be rescaled to have mean zero and standard deviation 1/2.  The authors showed that  these priors are weakly informative, not adding excess information on the analysis. 
\end{itemize} 
 We will complement the already done analysis by computing the WIM between the uniform and Cauchy priors on the bioassay data.  Besides the suggested scale parameter of the slope (2.5), we will also take 5 and 10 as scale parameters as in \cite{RMN2014} who investigated a similar problem for a distinct dataset. All models were fitted in $\mathtt{R}$ using the STAN language \citep{Bru2018} with the NUTS sampler (4 chains of length 2000, burn-in 1000, thinning 3). Figures~\ref{Fig314} and~\ref{Fig315} plot the posteriors for the regression coefficients and the LD50 parameter, respectively.

Table~\ref{VWdata}  presents the WIM for the bioassay dataset.  For all Cauchy-prior based posteriors, as much as the scale parameter of the slope's prior increases, the joint and marginal Wasserstein distances with the posterior based on the uniform prior decrease, which is not surprising since a larger spread of the Cauchy distribution implies more and more uniform-like behavior. This conclusion of decreasing WIM holds also for the quantity of interest LD50. We further notice the dependency between both marginals since changing only the slope's scale impacts also the intercept's WIM. A  look at the marginals reveals  that  the distance between joint posteriors seems mainly guided by the slope whose marginal WIM is quite close  to the WIM of the joint distribution. This however is likely due to the larger values of $\beta_1$ compared to $\beta_0$; though smaller in absolute values, the WIM for $\beta_0$ varies more in percentage than the WIM for $\beta_1$. We thus recommend that in future studies the judgement of the prior impact should be done on the marginal posteriors to get a full picture.

 \begin{table}
\begin{center}
\begin{tabular}{|l c c c c|}
\hline
  & $(\beta_0,\beta_1)$ & $\beta_0$ & $\beta_1$ & LD50 \\
\hline
Uniform \mbox{vs.} Cauchy(0, 2.5) &6.115 & 0.129 & 6.113 & 0.028 \\
 Uniform \mbox{vs.}                  Cauchy(0, 5.0)&5.162 & 0.090 & 5.161 & 0.017\\
 Uniform \mbox{vs.}                 Cauchy(0, 10.0) & 3.851 & 0.060 & 3.850 & 0.013\\
\hline
\end{tabular}
\caption{WIM between the posteriors (joint, marginal and LD50) based on the uniform prior and the weakly-informative Cauchy prior for the bioassay data. Only the prior for $\beta_1$ is indicated in the Cauchy settings, since for $\beta_0$ we always use Cauchy(0,10).}\label{VWdata}
\end{center}
\end{table}

\section{Discussion}\label{sec:discu}

In this paper we have introduced the WIM, a practically oriented measure of prior impact that allows comparing any two priors by quantifying the Wasserstein distance between the resulting posteriors. Our proposal thus retains the appealing intuitive touch of the approach from \cite{LRS2017a} and \cite{GhaLey2019a} while palliating its drawbacks, meaning that the WIM can also deal with multi-parameter and multi-dimensional situations, with non-nested priors and complicated forms of the posteriors (which are becoming more and more routine in modern applications). Through a Monte Carlo simulation study we investigated the link between our WIM and the theoretical results from \cite{LRS2017a} and \cite{GhaLey2019a}. Via the same simulation examples we also compared our WIM to two prior impact measures from the literature, namely the concepts of Neutrality and MOPESS. We could see that the WIM is an attractive alternative to these known proposals, since in various cases it provides more information (e.g., when the Neutrality reaches its upper bound from a certain point on) and is better interpretable (thanks to a monotone prior measure, which both the Neutrality and the MOPESS can lack). Moreover, it does not suffer from the high variability that the MOPESS exhibits. 

We will now wrap up the comparison by discussing further properties. While the MOPESS is a comparative measure like the WIM, the Neutrality is an absolute measure. Unlike the WIM, the MOPESS requires choosing a baseline prior; an advantage of the MOPESS however is its sign which  allows finding out which of the two priors is closer \mbox{w.r.t.} the current data set. Our WIM is  rather quick to compute (as is Neutrality), unlike the MOPESS; indeed, for non-conjugate models  advanced sampling algorithms are necessary for the MOPESS and need to be repeated several times (in particular for higher dimensions), implying a long computation time. A further appealing property of our WIM is its universal usage. The two  different real data sets serve as proof of its 
 broad usage. They could not have been tackled via the two competitor impact measures. Indeed, the Neutrality cannot be applied on the frontier data since the maximum likelihood estimate lies at the boundary of the parameter space, a situation the Neutrality cannot handle by definition. Since no multivariate extension of the Neutrality exists in the literature, it can also not be used for the LD50 logistic regression problem. The MOPESS cannot be used there, either, because it would require to know the distribution for the covariate which is unrealistic. We did apply the MOPESS on the frontier dataset, however ran into the following  issue: the high variability of the MOPESS led to OPESS ranges $[5\% , 95\%]$ that are considerably wide and contain the zero value. This observation was also noted in  \cite{JTC2020} who  emphasized that care should be taken in interpreting the impact solely on basis of the MOPESS.    The aforementioned problems of the Neutrality and the MOPESS are all restrictions that the WIM does not possess.

\

\noindent ACKNOWLEDGMENTS

\noindent This research is supported by a BOF Starting Grant of Ghent University. The authors would further like to thank David E. Jones, Robert N. Trangucci and  Yang Chen for  helping them with the $\mathtt{R}$ code for the MOPESS.

\bibliographystyle{chicago}
\bibliography{RV_biblio}

\end{document}